\documentclass[preprint2]{proto}
\usepackage{natbib}
\bibpunct{(}{)}{;}{a}{,}{,}
\usepackage{times}

\def\lapprox{{_<\atop{^\sim}}}

\def\science{{Science}}
\def\natr{{Nature Reports}}
\def\pccp{{PCCP}}
\newenvironment{dcl_itemize}
{\begin{itemize}
  \setlength{\itemsep}{1pt}
  \setlength{\parskip}{0pt}
  \setlength{\parsep}{0pt}}
{\end{itemize}}

\begin{document}

\title{\textbf{\LARGE Water: from clouds to planets}}

\author {\textbf{\large Ewine F.\ van Dishoeck}} 
\affil{\small\em
  Leiden Observatory, Leiden University, The Netherlands; Max
  Planck Institute for Extraterrestrial Physics, Garching, Germany}

\author {\textbf{\large Edwin A.\ Bergin}}
\affil{\small\em University of Michigan, USA}

\author {\textbf{\large Dariusz C.\ Lis}}
\affil{\small\em California Institute of Technology, USA}

\author {\textbf{\large Jonathan I.\ Lunine}}
\affil{\small\em Cornell University, USA}

\begin{abstract}
\baselineskip = 11pt
\leftskip = 0.65in 
\rightskip = 0.65in
\parindent=1pc {\small 
  Results from recent space missions, in particular \emph{Spitzer}
  and \emph{Herschel}, have lead to significant progress in our
  understanding of the formation and transport of water from clouds to
  disks, planetesimals, and planets. In this review, we provide the
  underpinnings for the basic molecular physics and chemistry of water
and outline these advances in the context of water
  formation in space, its transport to a forming disk, its evolution
  in the disk, and finally the delivery to forming terrestrial worlds
  and accretion by gas giants. Throughout, we pay close attention to
  the disposition of water as vapor or solid and whether it might be
  subject to processing at any stage. The context of the water in the
  solar system and the isotopic ratios (D/H) in various bodies are
  discussed as grounding data point for this evolution. Additional
  advances include growing knowledge of the composition of atmospheres
  of extra-solar gas giants, which may be influenced by the variable
  phases of water in the protoplanetary disk. Further, the
  architecture of extra-solar systems leaves strong hints of dynamical
  interactions, which are important for the delivery of water and
subsequent evolution of
  planetary systems. We conclude with an exploration of water on Earth
  and note that all of the processes and key parameters identified
  here should also hold for exoplanetary systems.
  \\~\\~\\~}
\end{abstract}

\section{\textbf{INTRODUCTION}}

With nearly 1000 exoplanets discovered to date and statistics
indicating that every star hosts at least one planet
\citep{Batalha13}, the next step in our search for life elsewhere in
the universe is to characterize these planets.  The presence of water
on a planet is universally accepted as essential for its potential
habitability. Water in gaseous form acts as a coolant that allows
interstellar gas clouds to collapse to form stars, whereas water ice
facilitates the sticking of small dust particles that eventually must
grow to planetesimals and planets.  The development of life requires
liquid water and even the most primitive cellular life on Earth consists
primarily of water. Water assists many chemical reactions leading to
complexity by acting as an effective solvent. It shapes the geology
and climate on rocky planets, and is a major or primary constituent of
the solid bodies of the outer solar system.

How common are planets that contain water, and how does the water
content depend on the planet's formation history and other properties
of the star-planet system?  Thanks to a number of recent space
missions, culminating with the {\it Herschel Space Observatory}, an
enormous step forward has been made in our understanding of where
water is formed in space, what its abundance is in various physical
environments, and how it is transported from collapsing clouds to
forming planetary systems. At the same time, new results are emerging
on the water content of bodies in our own solar system and in the
atmospheres of known exoplanets.  This review attempts to synthesize
the results from these different fields by summarizing our current
understanding of the water trail from clouds to planets.

Speculations about the presence of water on Mars and other planets in
our solar system date back many centuries. Water is firmly detected as
gas in the atmospheres of all planets including Mercury and as ice on
the surfaces of the terrestrial planets, the Moon, several moons of
giant planets, asteroids, comets and Kuiper Belt Objects \citep[see
review by][]{Encrenaz08}. Evidence for past liquid water on Mars has
been strengthened by recent data from the Curiosity rover
\citep{Williams13}. Water has also been detected in spectra of the Sun
\citep{Wallace95} and those of other cool stars. In interstellar
space, gaseous water was detected more than 40 years ago in the Orion
nebula through its masing transition at 22 GHz
\citep[1 cm;][]{Cheung69} and water ice was discovered a few years
later through its infrared bands toward protostars
\citep{Gillett73}. Water vapor and ice have now been observed in many
star- and planet-forming regions throughout the galaxy \citep[reviews
by][]{Cernicharo05,Boogert08,Melnick09,Bergin12} and even in external
galaxies out to high redshifts
\citep[e.g.,][]{Shimonishi10,Lis11,Weiss13}.  Water is indeed
ubiquitous throughout the universe.

On their journey from clouds to cores, the water molecules encounter a
wide range of conditions, with temperatures ranging from $<$10 K in
cold prestellar cores to $\sim$2000 K in shocks and the inner regions
of protoplanetary disks. Densities vary from $\sim 10^4$
cm$^{-3}$ in molecular clouds to $10^{13}$ cm$^{-3}$ in the midplanes
of disks and $10^{19}$ cm$^{-3}$ in planetary atmospheres. The
chemistry naturally responds to these changing conditions. A major
question addressed here is to what extent the water molecules produced
in interstellar clouds are preserved all the way to exoplanetary
atmospheres, or whether water is produced in situ in planet-forming
regions. Understanding how, where and when water forms is critical for
answering the question whether water-containing planets are common.

\section{\bf H$_2$O PHYSICS AND CHEMISTRY}

This section reviews the basic physical and chemical properties of
water in its various forms, as relevant for interstellar and planetary
system conditions.  More details, examples and links to databases can
be found in the recent review by \citet{vanDishoeck13}.

\subsection{Water phases}
\label{sect:phases}

Water can exist as a gas (vapor or `steam'), as a solid (ice), or as
a liquid. At the low pressures of interstellar space, only water vapor
and ice occur, with the temperature at which the transition occurs
depending on density. At typical cloud densities of $10^4$ particles
cm$^{-3}$, water sublimates around 100~K \citep{Fraser01}, but at
densities of $10^{13}$ cm$^{-3}$,
corresponding to the midplanes of protoplanetary disks, the
sublimation temperature increases to $\sim$160~K. According to the
phase diagram of water, liquid water can exist above the triple point
at 273~K and 6.12~mbar ($\sim 10^{17}$ cm$^{-3}$). Such
pressures and temperatures are typically achieved at the surfaces of
bodies of the size of Mars or larger and at distances between 0.7 and
1.7 AU for a solar-type star. 

Water ice can take many different crystalline and amorphous forms
depending on temperature and pressure. At interstellar densities,
crystallization of an initially amorphous ice to the cubic
configuration, $I_c$, occurs around 90~K. This phase change is
irreversible: even when the ice is cooled down again, the crystal
structure remains and it therefore provides a record of the
temperature history of the ice. Below 90~K, interstellar ice is mostly
in a compact high-density amorphous (HDA) phase, which does not
naturally occur on planetary surfaces \citep{Jenniskens94}.  The
densities of water ice in the HDA, LDA and $I_c$ phases are 1.17, 0.94
and 0.92 gr cm$^{-3}$, respectively, much lower than those of rocks
(3.2--4.4 gr cm$^{-3}$ for magnesium-iron silicates).

Clathrate hydrates are crystalline water-based solids in which small
non-polar molecules can be trapped inside `cages' of the
hydrogen-bonded water molecules. 
They can be formed when a gas of water mixed with other species
condenses\footnote{Strictly speaking, the term condensation refers to
  the gas to liquid transition; we adopt here the astronomical parlance
  where it is also used to denote the gas-to-solid transition.} out at
high pressure and has enough entropy to form a stable clathrate
structure \citep{Lunine85,Mousis10}.  Clathrate hydrates are found in
large quantities on Earth, with methane clathrates on the deep ocean
floor and in permafrost as the best known examples.  They have been
postulated to occur in large quantities on other planets and icy solar
system bodies.

\subsection{Water spectroscopy}
\label{sect:spectroscopy}

Except for in-situ mass spectroscopy in planetary and cometary
atmospheres, all information about interstellar and solar system water
comes from spectroscopic data obtained with telescopes. Because of the
high abundance of water in the Earth's atmosphere, the bulk of the
data comes from space observatories.  Like any molecule, water has
electronic, vibrational and rotational energy levels.  Dipole-allowed
transitions between electronic states occur at ultraviolet (UV)
wavelengths, between vibrational states at near- to mid-infrared (IR)
wavelengths, and between rotational states from mid- to far-IR and
submillimeter wavelengths.

Interstellar water vapor observations target mostly the pure
rotational transitions. H$_2$O is an asymmetric rotor with a highly
irregular set of energy levels, characterized by quantum numbers
$J_{K_A K_C}$. Because water is a light molecule, the spacing of its
rotational energy levels is much larger than that of heavy rotors, such
as CO or CS, and the corresponding wavelengths much shorter (0.5 mm vs
3--7~mm for the lowest transitions). The nuclear spins of the two
hydrogen atoms can be either parallel or anti-parallel, and this
results in a grouping of the H$_2$O energy levels into ortho
($K_A+K_C=$odd) and para ($K_A+K_C=$even) ladders, with a statistical
weight ratio of 3:1, respectively. Radiative transitions between these
two ladders are forbidden to high order, and only chemical reactions
in which an H atom of water is exchanged with an H-atom of a reactant
can transform ortho- to para-H$_2$O and vice versa.

\begin{figure}[t]
\begin{centering}
\includegraphics[angle=0,width=0.35\textwidth]{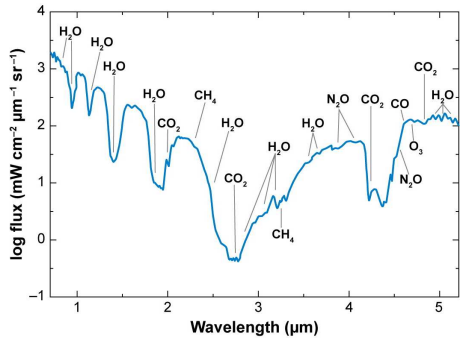}
\caption{The near-IR spectrum of the Earth showing the many water
  vibrational bands together with CO$_2$. The bands below 3 $\mu$m are
  due to overtones and combination bands and are often targeted in
  exoplanet searches. This spectrum was observed with the NIMS
  instrument on the Galileo spacecraft during its Earth flyby in
  December 1990. From \citet{Encrenaz08}, with permission from Annual
  Reviews, based on \citet{Drossart93}.}
\label{fig:earthspectrum}
\end{centering}
\end{figure}

Infrared spectroscopy can reveal both water vapor and ice.  Water has
three active vibrational modes: the fundamental $v$=1--0 bands of the
$\nu_1$ and $\nu_3$ symmetric and asymmetric stretches centered at 2.7
$\mu$m and 2.65 $\mu$m, respectively, and the $\nu_2$ bending mode at
6.2 $\mu$m. Overtone ($\Delta v=2$ or larger) and combination (e.g.,
$\nu_2$+$\nu_3$) transitions occur in hot gas at shorter wavelengths
(see Fig.~\ref{fig:earthspectrum} for example). Gas-phase water
therefore has a rich vibration-rotation spectrum with many individual
lines depending on the temperature of the gas. In contrast, the
vibrational bands of water ice have no rotational substructure and
consist of very broad profiles, with the much stronger $\nu_3$ band
overwhelming the weak $\nu_1$ band.  The ice profile shapes depend on
the morphology, temperature and environment of the water molecules
\citep{Hudgins93}. Crystalline water ice is readily distinguished by a
sharp feature around 3.1 $\mu$m that is lacking in amorphous water
ice.  Libration modes of crystalline water ice are found at 45 and 63
$\mu$m \citep{Moore94}.

Spectra of hydrous silicates (also known as phyllosilicates,
layer-lattice silicates or `clays') show sharp features at 2.70--2.75
$\mu$m due to isolated OH groups and a broader absorption from
2.75--3.2 $\mu$m caused by interlayered (`bound') water molecules.
 At longer wavelengths, various
peaks can occur depending on the composition; for example, the hydrous
silicate montmorillonite has bands at 49 and 100 $\mu$m
\citep{Koike82}.

Bound-bound electronic transitions of water occur at far-UV
wavelengths around 1240 \AA, but have not yet been detected in space.

\subsection{Water excitation}
\label{sect:excitation}
 
The strength of an emission or absorption line of water depends on the
number of molecules in the telescope beam and, for gaseous water, on
the populations of the individual energy levels.  These populations,
in turn, are determined by the balance between the collisional and
radiative excitation and de-excitation of the levels.  The radiative
processes involve both spontaneous emission and stimulated absorption
and emission by a radiation field produced by a nearby star, by warm
dust, or by the molecules themselves.

The main collisional partner in interstellar clouds is H$_2$. Accurate
state-to-state collisional rate coefficients, $C_{u\ell}$, of H$_2$O
with both ortho- and para-H$_2$ over a wide range of temperatures have
recently become available thanks to a dedicated chemical physics study
\citep{Daniel11}. Other collision partners such as H, He and electrons
are generally less important. In cometary atmospheres, water itself
provides most of the collisional excitation.

Astronomers traditionally analyze molecular observations through a
Boltzmann diagram, in which the level populations are plotted versus
the energy of the level involved.  The slope of the diagram gives the
inverse of the excitation temperature.
If collisional processes dominate over radiative
processes, the populations are in `local thermodynamic equilibrium'
(LTE) and the excitation temperature is equal to the kinetic
temperature of the gas, $T_{\rm ex}=T_{\rm kin}$. 
Generally level populations are far from LTE and
molecules are excited by collisions and de-excited by
spontaneous emission, leading to $T_{\rm ex}<T_{\rm kin}$. The critical density 
roughly delineates the transition between these regimes:
$n_{cr} = A_{u\ell}/C_{u\ell}$ and
therefore scales with $\mu_{u\ell}^2\nu_{u\ell}^3$, where $A$ is the Einstein
spontaneous emission coefficient, $\mu$ the electric dipole moment and $\nu$ the
frequency of the transition $u \to \ell$. In the case of water, the
combination of a large dipole moment (1.86 Debye) and high frequencies
results in high critical densities of $10^8$--$10^9$ cm$^{-3}$ for 
pure rotational transitions.

Analysis of water lines is much more complex than for simple molecules, such
as CO, for a variety of reasons. First, because of the large dipole
moment and high frequencies, the rotational transitions of water are
usually highly optically thick, even for abundances as low as $10^{-10}$.
Second, the water transitions
couple effectively with mid- and far-infrared radiation from warm
dust, which can pump higher energy levels. Third, the fact that the
`backbone' levels with $K_A$=0 or 1 have lower radiative decay rates
than higher $K_A$ levels can lead to population `inversion', in which
the population in the upper state divided by its statistical weight
exceeds that for the lower state (i.e., $T_{\rm ex}$ becomes
negative). Infrared pumping can also initiate this inversion. The
result is the well-known maser phenomenon, which is widely observed in
several water transitions in star-forming regions
\citep[e.g.,][]{Furuya03,Neufeld13,Hollenbach13}.
The bottom line is that accurate analysis of interstellar water
spectra often requires additional independent constraints, for example
from H$_2^{18}$O or H$_2^{17}$O isotopologues, whose abundances are
reduced by factors of about 550 and 2500, respectively, and whose
lines are more optically thin. At infrared wavelengths, lines are
often spectrally unresolved, which further hinders the interpretation.

\begin{figure}[t]
\begin{centering}
\includegraphics[angle=0,width=0.4\textwidth]{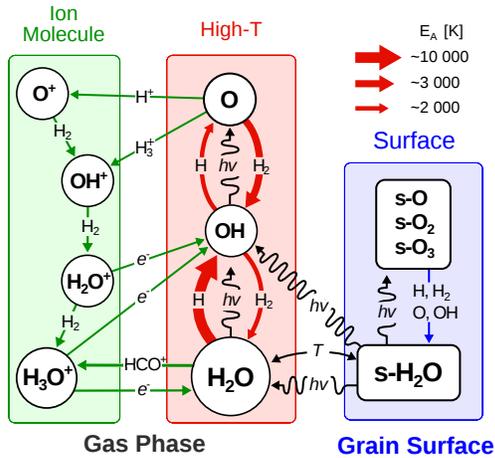}
\caption{Summary of the main gas-phase and solid-state chemical
  reactions leading to the formation and destruction of H$_2$O under
  non-equilibrium conditions.  Three different chemical regimes can be
  distinguished: (i) ion-molecule chemistry, which dominates gas-phase
  chemistry at low $T$; (ii) high-temperature neutral-neutral
  chemistry; and (iii) solid state chemistry.  $e$ stands for
  electron, $\nu$ for photon and $s-$X indicates that species X is on
  the grains. Simplified version of figure by \cite{vanDishoeck11}.}
\label{fig:network}
\end{centering}
\end{figure}

\subsection{Water chemistry}
\label{sect:chemistry}

\subsubsection{Elemental abundances and equilibrium chemistry}

The overall abundance of elemental oxygen with respect to total
hydrogen nuclei in the interstellar medium is estimated to be
$5.75\times 10^{-4}$ \citep{Przybilla08}, 
of which 16--24\% is locked
up in refractory silicate material in the diffuse interstellar medium
\citep{Whittet10}. The abundance of volatile oxygen (i.e., not tied up
in some refractory form) is measured to be $3.2\times 10^{-4}$ in
diffuse clouds \citep{Meyer98}, so this is the maximum amount of
oxygen that can cycle between water vapor and ice in dense
clouds. Counting up all the forms of detected oxygen in diffuse
clouds, the sum is less than the overall elemental oxygen
abudance. Thus, a fraction of oxygen is postulated to be in some yet unknown
refractory form, called UDO (`unknown depleted oxygen'), whose
fraction may increase from 20\% in diffuse clouds up to 50\% in
dense star-forming regions \citep{Whittet10}. For comparison, the
abundances of elemental carbon and nitrogen are $3\times 10^{-4}$ and
$1\times 10^{-4}$, respectively, with about 2/3 of the carbon thought
to be locked up in solid carbonaceous material.

For a gas in thermodynamic equilibrium (TE), the fractional abundance
of water is simply determined by the elemental composition of the gas
and the stabilities of the molecules and solids that can be produced
from it. For standard interstellar abundances\footnote{The notation
  [X] indicates the overall abundance 
  of element X in all forms, be it atoms, molecules or solids.} with
[O]/[C]$>1$, there
are two molecules in which oxygen can be locked up: CO and H$_2$O.  At
high pressures in TE, the fraction of CO results from the equilibrium
between CO and CH$_4$, with CO favored at higher temperatures.  For
the volatile elemental abundances quoted above, this results in an
H$_2$O fractional abundance of $(2-3)\times 10^{-4}$ with respect to
total hydrogen, if the CO fractional abundance ranges from $(0-1)\times
10^{-4}$. With respect to H$_2$, the water abundance would then be
$(5-6)\times 10^{-4}$ assuming that the fraction of hydrogen in atomic
form is negligible (the density of hydrogen nuclei $n_{\rm H}=n$(H) +
2$n$(H$_2$)).  Equilibrium chemistry is established at densities above
roughly $10^{13}$ cm$^{-3}$, when three body processes become
significant.  Such conditions are found in planetary atmospheres and
in the shielded midplanes of the inner few AU of protoplanetary disks.

Under most conditions in interstellar space, however, the densities
are too low for equilibrium chemistry to be established. Also, strong
UV irradiation drives the chemistry out of equilibrium, even in 
high-density environments, such as the upper atmospheres of planets and
disks. Under these conditions, the fractional abundances are
determined by the kinetics of the two-body reactions between the
various species in the gas. Figure~\ref{fig:network} summarizes the
three routes to water formation that have been identified. Each of
these routes dominates in a specific environment.

\subsubsection{Low temperature gas-phase chemistry}
\label{sect:hight}

In diffuse and translucent interstellar clouds with densities less than $\sim 10^4$
cm$^{-3}$ and temperatures below 100 K, water is formed largely by a
series of ion-molecule reactions
\citep[e.g.,][]{Herbst73}. 
  The network starts with the reactions
  O + H$_3^+$ and O$^+$ + H$_2$ leading to OH$^+$.
The H$_3^+$ ion is produced by interactions of energetic
cosmic-ray particles with the gas, producing H$_2^+$ and H$^+$, with
the subsequent fast reaction of H$_2^+$ + H$_2$ leading to H$_3^+$.
The cosmic ray ionization rate of atomic hydrogen denoted by
$\zeta_{\rm H}$ can be as high as $10^{-15}$ s$^{-1}$ in some diffuse
clouds, but drops to $10^{-17}$ s$^{-1}$ in denser regions
\citep{Indriolo12,Rimmer12}.  The ionization rate of H$_2$ is
$\zeta_{\rm H_2}\approx 2 \zeta_{\rm H}$.

A series of rapid reactions of OH$^+$ and
H$_2$O$^+$ with H$_2$ lead to H$_3$O$^+$, which can dissociatively
recombine to form H$_2$O and OH with branching ratios of $\sim$0.17
and 0.83, respectively \citep{Buhr10}. H$_2$O is destroyed by
photodissociation and by reactions with C$^+$, H$_3^+$ and other ions
such as HCO$^+$. Photodissociation of H$_2$O starts to be effective
shortward of 1800 \AA \ and continues down to the ionization threshold
at 983 \AA \ (12.61 eV), including Ly $\alpha$ at 1216 \AA. Its
lifetime in the general interstellar radiation field, as given by
\citet{Draine78}, is only 40~yr.

\subsubsection{High-temperature gas-phase chemistry}

At temperatures above 230~K, the energy barriers for reactions with
H$_2$ can be overcome and the reaction O + H$_2$ $\to$ OH + H becomes
the dominant channel initiating water formation \citep{Elitzur78}. OH
subsequently reacts with H$_2$ to form H$_2$O, a reaction which is
exothermic, but has an energy barrier of $\sim$2100~K
\citep{Atkinson04}.  This route drives all the available gas-phase
oxygen into H$_2$O, unless strong UV or a high atomic H abundance
convert some water back to OH and O.  High-temperature chemistry
dominates the formation of water in shocks, in the inner envelopes
around protostars, and in the warm surface layers of protoplanetary
disks.

\subsubsection{Ice chemistry}
\label{sect:icechemistry}

The timescale for an atom or molecule to collide with a grain and
stick to it is $t_{\rm fo}=3\times 10^9/n_{\rm H_2}$ yr for normal size grains
and sticking probabilities close to unity \citep{Hollenbach09}.
Thus, for densities greater than $10^4$ cm$^{-3}$, the time scales for
freeze-out are less than a few $\times 10^5$ yr, generally smaller
than the lifetime of dense cores (at least $10^5$ yr). Reactions
involving dust grains are therefore an integral part of the
chemistry. Even weakly bound species, such as atomic H, have a long
enough residence time on the grains at temperatures of 10--20 K to
react; H$_2$ also participates in some surface reactions, but remains
largely in the gas.  \citet{Tielens82} postulated that the formation
of water from O atoms proceeds through three routes involving
hydrogenation of $s$-O, $s$-O$_2$ and $s$-O$_3$, respectively, where
$s$-X indicates a species on the surface. All three routes have
recently been verified and quantified in the laboratory and detailed
networks with simulations have been drawn up \citep[see][for
summaries]{Cuppen10,Oba12,Lamberts13}.

Water ice formation is in competition with various desorption
processes, which limit the ice build-up. At dust temperatures below
the thermal sublimation limit, photodesorption is an effective
mechanism to get species back to the gas phase, although only a small
fraction of the UV absorptions results in desorption of intact H$_2$O
molecules \citep{Andersson08}.
The efficiency is about
$10^{-3}$ per incident photon, as determined through laboratory
experiments and theory \citep{Westley95,Oberg09h2o,Arasa10}.  Only the
top few monolayers of the ice contribute. 
The UV needed to trigger photodesorption can come either
from a nearby star, or from the general interstellar radiation
field. Deep inside clouds, cosmic rays produce a low level of UV flux,
$\sim 10^4$ photons cm$^{-2}$ s$^{-1}$, through interaction with H$_2$
\citep{Prasad83}.  Photodesorption via X-rays is judged to be
inefficient, although there are large uncertainties in the
transfer of heat within a porous aggregate \citep[][]{Najita01}.  UV
photodesorption of ice is thought to dominate the production of
gaseous water in cold pre-stellar cores, the cold outer envelopes
of protostars and the outer parts of protoplanetary disks.

Other non-thermal ice desorption processes include cosmic ray induced
spot heating (which works for CO, but is generally not efficient for
strongly bound molecules like H$_2$O) and desorption due to the energy
liberated by the reaction (called `reactive' or `chemical'
desorption). These processes are less well explored than
photodesorption, but a recent laboratory study of $s$-D + $s$-OD $\to$
$s$-D$_2$O suggests that as much as 90\% of the product can be
released into the gas phase \citep{Dulieu13}. The details of this
mechanism, which has not yet been included in models, are not yet
understood and may strongly depend on the substrate.

Once the dust temperature rises above $\sim$100~K (precise value
being pressure dependent), H$_2$O ice thermally sublimates on timescales of
years, leading to initial gas-phase abundances of H$_2$O as high as
the original ice abundances.
These simulations use
a binding energy of 5600 K for amorphous ice and a slightly higher
value of 5770 K for crystalline ice, derived from laboratory
experiments \citep{Fraser01}.  Thermal desorption of ices contributes
to the gas-phase water abundance in the warm inner protostellar
envelopes (`hot cores') and inside the snow line in 
disks.

\subsubsection{Water deuteration}
\label{sect:deuteration}

Deuterated water, HDO and D$_2$O, is formed through the same processes
as illustrated in Figure~\ref{fig:network}. There are, however, a number
of chemical processes that can enhance the HDO/H$_2$O and
D$_2$O/H$_2$O ratios by orders of magnitude compared with the overall
[D]/[H] ratio of $2.0\times 10^{-5}$ in the local interstellar medium
\citep{Prodanovic10}. A detailed description is
given in the chapter by \emph{Ceccarelli et al.}, here only a brief summary
is provided.

In terms of pure gas-phase chemistry, the direct exchange reaction
H$_2$O +HD $\leftrightarrow$ HDO $+$ H$_2$ is often considered in
solar system models \citep{Richet77}. In thermochemical equilibrium
this reaction can provide at most a factor of 3 enhancement, and even
that may be limited by kinetics \citep{Lecluse94}.  The exchange
reaction D + OH $\to$ H + OD, which has a barrier of $\sim$100~K
\citep{Sultanov04}, is particularly effective in high-temperature gas
such as present in the inner disk \citep{Thi10dh}.
Photodissociation of HDO
enhances OD compared with OH by a factor of 2--3, which could be a
route to further fractionation.

The bulk of the deuterium fractionation in cold clouds comes from
gas-grain processes.  \citet{Tielens83} pointed out that the fraction
of deuterium relative to hydrogen atoms arriving on a grain surface,
D/H, is much higher than the overall [D]/[H] ratio, which can be
implanted into molecules in the ice. This naturally leads to enhanced
formation of OD, HDO and D$_2$O ice according to the grain-surface
formation routes.  The high atomic D/H ratio in the gas arises from
the enhanced gaseous H$_2$D$^+$, HD$_2^+$, and D$_3^+$ abundances at
low temperatures ($\leq$25 K), when the ortho-H$_2$ abundance drops
and their main destroyer, CO, freezes out on the grains
\citep{Pagani92,Roberts03}. Dissociative recombination with electrons
then produces enhanced D.  The enhanced H$_2$D$^+$ also leads to
enhanced H$_2$DO$^+$ and thus HDO in cold gas, but this is usually a
minor route compared with gas-grain processes.

On the grains, tunneling reactions can have the opposite effect,
reducing the deuterium fractionation. For example, the OD + H$_2$
tunneling reaction producing HDO ice is expected to occur slower than
the OH + H$_2$ reaction leading to H$_2$O ice. On the other hand,
thermal exchange reactions in the ice, such as H$_2$O + OD $\to$ HDO +
OH have been shown to occur rapidly in ices at higher temperatures;
these can both enhance and decrease the fractionation.  
Both thermal desorption at high ice temperatures and photodesorption
at low ice temperatures have a negligible effect on the deuterium
fractionation, i.e., the gaseous HDO/H$_2$O and D$_2$O/H$_2$O ratios
reflect the ice ratios if no other gas-phase processes are involved.

\section{\bf CLOUDS AND PRE-STELLAR CORES: ONSET OF WATER FORMATION}
\label{sect:clouds}

In this and following sections, our knowledge of the water reservoirs
during the various evolutionary stages from clouds to planets will be
discussed. The focus is on low-mass protostars ($<$100 L$_\odot$) and
pre-main sequence stars (spectral type A or later). Unless stated
otherwise, fractional abundances are quoted with respect to H$_2$ and
are simply called `abundances'. Often the denominator, i.e., the
(column) density of H$_2$, is more uncertain than the numerator.

The bulk of the water in space is formed on the surfaces of dust
grains in dense molecular clouds. Although a small amount of water is
produced in the gas in diffuse molecular clouds through 
ion-molecule chemistry, its abundance of $\sim 10^{-8}$ found by {\it
  Herschel}-HIFI  \citep{Flagey13} is negligible compared
with that produced in the solid state. In contrast, observations of
the 3 $\mu$m water ice band toward numerous infrared sources behind
molecular clouds, from the ground and from space, show that water ice
formation starts at a threshold extinction of $A_V\approx 3$~mag
\citep{Whittet13}. 
These clouds have
densities of at least 1000~cm$^{-3}$, but are not yet collapsing to
form stars. The ice abundance is $s$-H$_2$O/H$_2 \approx
5\times 10^{-5}$, indicating that a significant fraction of the
available oxygen has been transformed to water ice even at this early
stage \citep{Whittet88,Murakawa00,Boogert11}.  Such high ice
abundances are too large to result from freeze-out of gas-phase water
produced by ion-molecule reactions.

\begin{figure}[t]
\begin{centering}
  \includegraphics[width=0.35\textwidth]{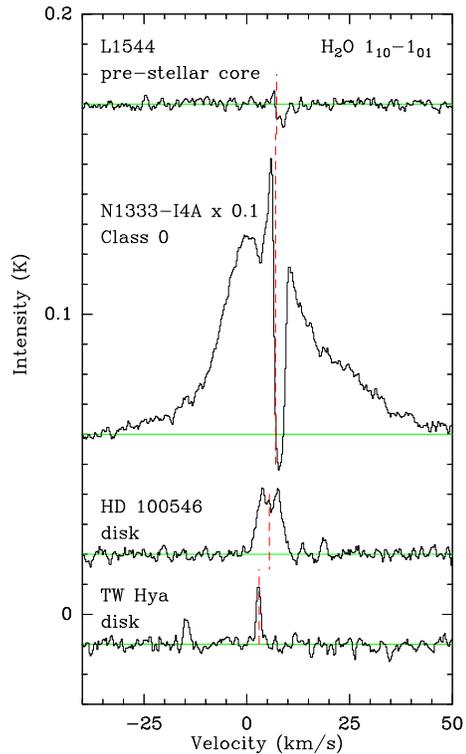}
  \caption{{\it Herschel}-HIFI spectra of the H$_2$O
    $1_{10}$--$1_{01}$ line at 557 GHz in a pre-stellar core (top),
    protostellar envelope (middle) and two protoplanetary disks
    (bottom) (spectra shifted vertically for clarity). The red dashed
    line indicates the rest velocity of the source. Note the different
    scales: water vapor emission is strong toward protostars, but very weak
    in cold cores and disks. The feature at -15 km s$^{-1}$ in the TW
    Hya spectrum is due to NH$_3$. Figure by L. Kristensen, adapted
    from \citet{Caselli12}, \citet{Kristensen12} and \citet[][and in
    prep.]{Hogerheijde11}. }
  \label{fig:557spectra}
\end{centering}
\end{figure}

The densest cold cores just prior to collapse have such high
extinctions that direct IR ice observations are not possible. In
contrast, the water reservoir (gas plus ice) can be inferred from {\it
  Herschel}-HIFI observations of such cores.
Fig.~\ref{fig:557spectra} presents the detection of the H$_2$O
$1_{10}$--$1_{01}$ 557 GHz line toward L1544 \citep{Caselli12}. The
line shows blue-shifted emission and red-shifted absorption,
indicative of inward motions in the core. Because of the high critical
density of water, the emission indicates that water vapor must be
present in the dense central part. The infalling red-shifted gas
originates on the near-side.  Because the different parts of the line
profile probe different parts of the core, the line shape can be used
to reconstruct the water vapor abundance as a function of position
throughout the entire core.

The best-fit water abundance profile is obtained with a simple
gas-grain model, in which atomic O is converted into water ice on the
grains, with only a small fraction returned back into the gas by
photodesorption \citep{Bergin00,Roberts02,Hollenbach09}.  The maximum
gas-phase water abundance of $\sim 10^{-7}$ occurs in a ring at the
edge of the core around $A_V\approx 4$ mag, where external UV photons
can still penetrate to photodesorb the ice, but where they are no
longer effective in photodissociating the water vapor.  In the central
shielded part of the core, cosmic ray induced UV photons keep a small,
$\sim 10^{-9}$, but measurable fraction of water in the gas
\citep{Caselli12}.  Quantitatively, the models indicate that the bulk
of the available oxygen has been transformed into water ice in the
core, with an ice abundance of $\sim 10^{-4}$ with respect to H$_2$.

\section{\bf PROTOSTARS AND OUTFLOWS}

\subsection{Outflows}
\label{sect:outflow}

{\it Herschel}-HIFI and PACS data show strong and broad water
profiles characteristic of shocks associated with embedded protostars,
from low to high mass. In fact, for low-mass protostars this shocked
water emission completely overwhelms the narrower lines from the bulk
of the collapsing envelope, even though the shocks contain less than
1\% of the mass of the system. 
Maps of the water emission around solar-mass protostars such as
L1157 reveal water not only at the protostellar
position but also along the outflow at `hot spots' where the
precessing jet interacts with the cloud \citep{Nisini10}. Thus, water
traces the {\it currently} shocked gas at positions, which are somewhat
offset from the bulk of the cooler entrained outflow gas seen in the
red- and blue-shifted lobes of low-$J$ CO lines \citep{Tafalla13,LeFloch10}.

Determinations of the water abundance in shocks vary from values as
low as $10^{-7}$ to as high as $10^{-4}$ \citep[see][for
summary]{vanDishoeck13}.  In non-dissociative shocks, the
temperature reaches values of a few thousand K and all available
oxygen is expected to be driven into water \citep{Kaufman96}.  The low
values likely point to the importance of UV radiation in the shock
chemistry and shock structure. For the purposes of this chapter, the
main point is that even though water is rapidly produced in shocks at
potentially high abundances, the amount of water contained in the
shocks is small, and, moreover, most of it is lost to space through outflows.

\begin{figure}[t]
\begin{centering}
\includegraphics[angle=0,width=0.4\textwidth]{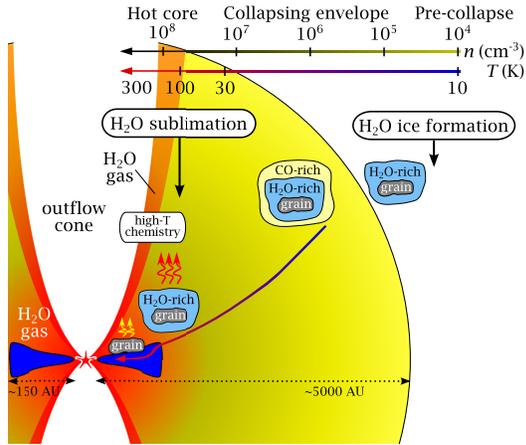}
\caption{Schematic representation of a protostellar envelope and
  embedded disk with key steps in the water chemistry indicated. Water
  ice is formed in the parent cloud before collapse and stays mostly
  as ice until the ice sublimation temperature of $\sim$100 K close to
  the protostar is reached. Hot water is formed in high abundances in
  shocks associated with the outflow, but this water is not
  incorporated into the planet-forming disk. Figure by R. Visser,
  adapted from \citet{Herbst09}.}
\label{fig:cartoon}
\end{centering}
\end{figure}

\subsection{Protostellar envelopes: the cold outer reservoir}
\label{sect:protostar}

As the cloud collapses to form a protostar in the center, the
water-ice coated grains created in the natal molecular cloud move
inward, feeding the growing star and its surrounding disk
(Fig.~\ref{fig:cartoon}). The water ice abundance can be measured
directly through infrared spectroscopy of various water ice bands
toward the protostar itself. Close to a hundred sources have been
observed, from very low luminosity objects (`proto-brown dwarfs') to
the highest mass protostars
\citep{Gibb04,Pontoppidan04,Boogert08,Zasowski09,Oberg11}.  Inferred
ice abundances with respect to H$_2$ integrated along the line of
sight are (0.5--1)$\times$$10^{-4}$.

The water vapor abundance in protostellar envelopes is probed through
spectrally-resolved {\it Herschel}-HIFI lines.  Because the gaseous
water line profiles are dominated by broad outflow emission
(Fig.~\ref{fig:557spectra}), this component needs to be subtracted, or
an optically thin water isotopologue needs to be used to determine the
quiescent water.  
Clues to the water vapor abundance structure can be
obtained through narrow absorption and emission
features in so-called (inverse) P-Cygni profiles
(see NGC 1333 IRAS 4A in Fig.~\ref{fig:557spectra}).
The analysis of these data proceeds along the same lines
as for pre-stellar cores. The main difference is that the dust
temperature now increases inwards, from a
low value of 10--20 K at the edge to a high value of several hundred K
in the center of the core (Fig.~\ref{fig:cartoon}). In the simplest
spherically symmetric case, the density follows a power-law 
$n\propto r^{-p}$ with $p$=1--2.  
As for pre-stellar cores, the data require the
presence of a photodesorption layer at the edge of the core with a
decreasing water abundance at smaller radii, where gaseous water is
maintained by the cosmic ray induced photodesorption of water ice
\citep{Coutens12,Mottram13}.  Analysis of the combined gaseous water
and water ice data for the same source shows that the ice/gas ratio is
at least $10^4$ \citep[][]{Boonman03h2o}. Thus, the bulk of the water
stays in the ice in this cold part, at a high abundance of $\sim
10^{-4}$ as indicated by direct measurements of both the water ice and
gas.

\subsection{Protostellar envelopes: the warm inner part}
\label{sect:hotcore}

When the infalling parcel enters the radius at which the dust
temperature reaches $\sim$100 K, the gaseous water abundance jumps
from a low value around $10^{-10}$ to values as high as $10^{-4}$
\citep[e.g.,][]{Boonman03,Herpin12,Coutens12}.  The 100 K
radius scales roughly as $2.3\times 10^{14} \sqrt(L/L_\odot)$ cm
\citep{Bisschop07}, and is small, $<$100 AU, for low-mass sources and a few
thousand AU for high-mass protostars. The precise abundance of water
in the warm gas is still uncertain, however, and can range from
$10^{-6}$--$10^{-4}$ depending on the source and analysis
\citep{Emprechtinger13,Visser13}. A high water abundance would
indicate that all water sublimates from the grains in the `hot core'
before the material enters the disk; a low abundance the opposite.

The fate of water in protostellar envelopes on scales of the size of
the embedded disk is currently not well understood, yet it is a
crucial step in the water trail from clouds to disks. 
To probe the inner few hundred AU, a high excitation line of a water
isotopolog line not dominated by the outflow or high angular
resolution is needed: ground-based millimeter interferometry of the
H$_2^{18}$O $3_{13}-2_{20}$ ($E_u=204$ K) line at 203 GHz
\citep{Persson12} and deep {\it Herschel}-HIFI spectra of excited
H$_2^{18}$O or H$_2^{17}$O lines, such as the $3_{12}-3_{03}$ ($E_u=249$ K) line at
1095~GHz have been used. 
Two main problems need to be faced in the analysis. First, comparison
of ground-based and {\it Herschel} lines for the same source show that
the high frequency HIFI lines can be optically thick even for
H$_2$$^{18}$O and H$_2^{17}$O, because of their much higher Einstein
$A$ coefficients. Second, the physical structure of the envelope and
embedded disk on scales of a few hundred AU is not well understood
\citep{Jorgensen05i2}, so that abundances are difficult to
determine since the column of warm H$_2$ is poorly
constrained. Compact flattened dust structures are not necessarily
disks in Keplerian rotation \citep{Chiang08} and only a fraction of
this material may be at high temperatures.

\citet{Jorgensen10} and \citet{Persson12} measure water columns and
use H$_2$ columns derived from continuum interferometry data on the
same scales ($\sim$1$''$) to determine water abundances of $\sim 10^{-8}
- 10^{-5}$ for three low-mass protostars,
consistent with the fact that the bulk of the gas on these scales is
cold and water is frozen. From a combined analysis of the
interferometric and HIFI data, using C$^{18}$O 9--8 and 10--9 data to
determine the {\it warm} H$_2$ column, \citet{Visser13} infer water
abundances of $2 \times 10^{-5} - 2\times 10^{-4}$ in the $\geq$100 K
gas, as expected for the larger-scale hot cores.

The important implication of these results is that the bulk of the
water stays as ice in the inner few hundred AU and that only a few \%
of the dust may be at high enough temperatures to thermally sublimate
H$_2$O .  This small fraction of gas passing through high-temperature
conditions for ice sublimation is consistent with 2D models of
collapsing envelope and disk formation, which give fractions
of $<1-20$\% depending on initial conditions
\citep{Visser09,Visser11,Ilee11,Harsono13,Hincelin13}.

\subsection{Entering the disk: the accretion shock and history of water in disks}
\label{sect:accretion}

The fact that only a small fraction of the material within a few hundred AU
radius is at $\geq$ 100 K (\S~\ref{sect:hotcore})
implies that most of the water is present as ice and is still moving
inwards (Fig.~\ref{fig:cartoon}). At some radius, however, the
high-velocity infalling parcels must encounter the low-velocity
embedded disk, resulting in a shock at the boundary. This shock
results in higher dust temperatures behind the shock front than those
achieved by stellar heating (\citealt{Neufeld94}; see
\citealt{Visser09} for a simple fitting formula) and can also sputter
ices. At early times, accretion velocities are high and all ices would
sublimate or experience a shock strong enough to induce
sputtering. However, this material normally ends up in the star rather
than in the disk, so it is not of interest for the current story. The
bulk of the disk is thought to be made up through layered accretion of
parcels that fall in later in the collapse process, and which enter
the disk at large radii, where the shock is much weaker
\citep{Visser09}. Indeed, the narrow line widths of H$_2^{18}$O of
only 1 km s$^{-1}$ seen in the interferometric data
\citep{Jorgensen10} argue against earlier suggestions, based on {\it
  Spitzer} data, of large amounts of hot water going through an
accretion shock in the embedded phase, or even being created through
high-temperature chemistry in such a shock \citep{Watson07}. This view
that accretion shocks do not play a role also contrasts with the
traditional view in the solar system community that all ices evaporate
and recondense when entering the disk \citep{Lunine91,Owen93}.

\begin{figure}[t]
\begin{centering}
\includegraphics[angle=0,width=0.50\textwidth]{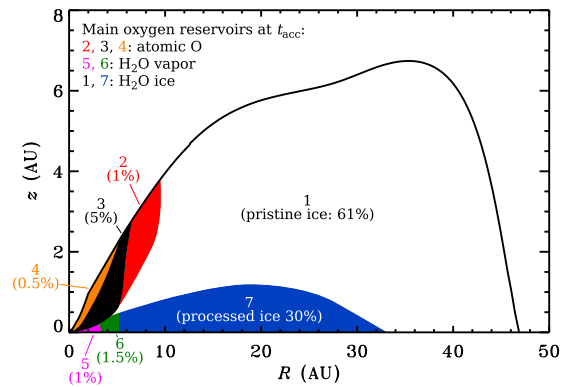}
\caption{Schematic view of the history of H$_2$O gas and ice
  throughout a young disk at the end of the accretion phase. The main
  oxygen reservoir is indicated for each zone. The percentages indicate the
  fraction of disk mass contained in each zone. Zone~1 contains
  pristine H$_2$O formed prior to star formation and never altered
  during the trajectory from cloud to disk. In Zone 7, the ice has
  sublimated once and recondensed again.  Thus, the ice in planet- and
  planetesimal-forming zones of disks is a mix of pristine and
  processed ice.  From \citet{Visser11}.}
\label{fig:diskhistory}
\end{centering}
\end{figure}

Figure~\ref{fig:diskhistory} shows the history of water molecules in
disks at the end of the collapse phase at $t_{\rm acc}=2.5\times 10^5$
yr for a standard model with an initial core mass of 1 M$_\odot$,
angular momentum $\Omega_0=10^{-14}$~s$^{-1}$ and sound speed
$c_s=0.26$ km s$^{-1}$ \citep{Visser11}. The material ending up in
zone 1 is the only water that is completely `pristine', i.e.,
formed as ice in the cloud and never sublimated, ending up
intact in the disk. Material ending up in the other zones contains
water that sublimated at some point along the infalling trajectory. In
zones 2, 3 and 4, close to the outflow cavity, most of the oxygen is
in atomic form due to photodissociation, with varying degrees of
subsequent reformation. In zones 5 and 6, most oxygen is in gaseous
water.
Material in zone 7 enters the disk early
and comes close enough to the star to sublimate. This material does not
end up in the star, however, but is transported outward in the disk
to conserve angular momentum, re-freezing when the temperature becomes
low enough. The detailed chemistry and fractions of water in each of
these zones depend on the adopted physical model and on whether
vertical mixing is included \citep{Semenov11}, but the
overall picture is robust.

\section{\bf PROTOPLANETARY DISKS}

Once accretion stops and the envelope has dissipated, a pre-main
sequence star is left, surrounded by a disk of gas and dust. These
protoplanetary disks form the crucial link between material in clouds
and that in planetary systems.  Thanks to the new observational
facilities, combined with sophisticated disk
chemistry models, the various water reservoirs in disks are now
starting to be mapped out. Throughout this chapter, we will call the
disk out of which our own solar system formed the `solar nebula disk'.
\footnote{Alternative nomenclatures in the literature include
  `primordial disk', `presolar disk', `protosolar nebula' or
  `primitive solar nebula'.}

\subsection{Hot and cold water in disks: observations}
\label{sect:diskobs}

With increasing wavelengths, regions further out and deeper into the
disk can be probed. 
The surface layers of the inner few AU of disks are probed by near-
and mid-IR observations. {\it Spitzer}-IRS detected a surprising wealth of
highly-excited pure rotational
lines of warm water at 10--30 $\mu$m 
\citep{Carr08,Salyk08}, and these lines have since been shown to be
ubiquitous in disks around low-mass T Tauri stars
\citep{Pontoppidan10,Salyk11}, with line profiles consistent with a
disk origin \citep{Pontoppidan10visir}.  Typical water excitation
temperatures are $T_{\rm ex}$$\approx$450 K. Spectrally resolved
ground-based near-IR vibration-rotation lines around 3 $\mu$m show
that in some sources the water originates in both a disk and a slow
disk wind \citep{Salyk08,Mandell12}. Abundance ratios are difficult to
extract from the observations, because the lines are highly saturated
and, in the case of {\it Spitzer} data, spectrally unresolved. Also,
the IR lines only probe down to moderate height in the disk until the
dust becomes optically thick.  Nevertheless, within the more than an
order of magnitude uncertainty, abundance ratios of
H$_2$O/CO$\sim$1--10 have been inferred for emitting radii up to a few
AU \citep{Salyk11,Mandell12}. This indicates that the inner disks have
high water abundances of order $\sim 10^{-4}$ and are thus not dry, at
least not in their surface layers.  The IR data show a clear dichotomy
in H$_2$O detection rate between disks around the lower-mass T Tauri
stars and higher-mass, hotter A-type stars
\citep{Pontoppidan10,Fedele11}. Also, transition disks with inner dust
holes show a lack of water line emission. This is likely due to 
more rapid photodissociation by stars with higher $T_*$,
and thus stronger UV radiation, in regions where the molecules are
not shielded by dust.

\begin{figure}[t]
\begin{centering}
\includegraphics[angle=0,width=0.44\textwidth]{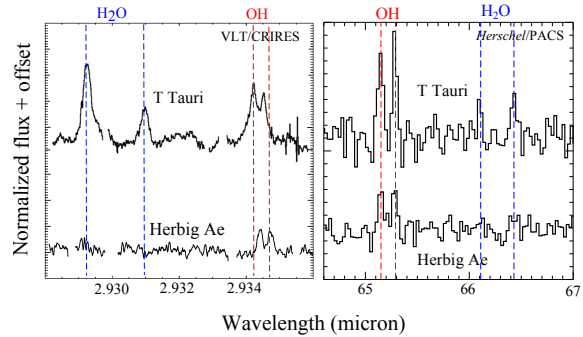}
\caption{Near-IR (left) and far-IR (right) spectra of a T Tau and a
  Herbig Ae disk, showing OH lines in both but H$_2$O primarily in
  disks around cooler T Tau stars. Figure by D.\ Fedele, based on
  \citet{Fedele11,Fedele13}.}
\label{fig:fedele}
\end{centering}
\end{figure}

Moving to longer wavelengths, {\it Herschel}-PACS spectra probe gas at
intermediate radii of the disk, out to 100 AU. Far-IR lines from warm
water have been detected in a few disks
\citep{Riviere12,Meeus12,Fedele12,Fedele13}. As for the inner disk,
the abundance ratios derived from these data are highly uncertain.
Sources in which both H$_2$O and CO far-infrared lines have been
detected (only a few) indicate H$_2$O/CO column density ratios of
$10^{-1}$, suggesting a water abundance of order $10^{-5}$ at
intermediate layers, but upper limits in other disks suggest values
that may be significantly less. Again the disks around T Tauri stars
appear to be richer in water than those around A-type stars
(Fig.~\ref{fig:fedele}).

In principle, the pattern of water lines with wavelength should allow
the transition from the gaseous water-rich to the water-poor (the snow
line) to be probed. As shown by LTE excitation disk models, the
largest sensitivity to the location of the snow line is provided by
lines in the 40--60 $\mu$m region, which is exactly the wavelength
range without observational facilities except for SOFIA
\citep{Meijerink09}.  For one disk, that around TW Hya, the available
shorter and longer wavelength water data have been used to put
together a water abundance profile across the entire disk
\citep{Zhang13}. This disk has a dust hole within 4~AU, within which
water is found to be depleted. The water abundance rises sharply to a
high abundance at the inner edge of the outer disk at 4~AU, but then
drops again to very low values as water freezes out in the cold outer
disk.

The cold gaseous water reservoir beyond 100 AU is uniquely probed by
{\it Herschel}-HIFI data of the ground rotational transitions at 557
and 1113 GHz. Weak, but clear detections of both lines have been
obtained in two disks, around the nearby T Tau star TW Hya
\citep{Hogerheijde11} and the Herbig Ae star HD 100546
(\emph{Hogerheijde et al.}, in prep.)
(Fig.~\ref{fig:557spectra}). These are the deepest integrations
obtained with the HIFI instrument, with integration times up to 25~hr
per line.
Similarly deep integrations on 5 other disks do not show
detections of water at the same level, nor do shallower observations
of a dozen other disks of different characteristics. One possible
exception, DG Tau \citep{Podio13}, is a late class I source with a
well-known jet and a high X-ray flux. 
The TW Hya detection implies abundances of gaseous water around
$10^{-7}$ in the intermediate layer of the disk, with the bulk of the
oxygen in ice on grains at lower layers. Quantitatively, 0.005 Earth
oceans of gaseous water and a few thousand oceans of water ice have
been detected (1 Earth ocean = 1.4$\times 10^{24}$ gr=0.00023 M$_{\rm
  Earth}$). While this is plenty of water to seed an Earth-like planet
with water, a single Jovian-type planet formed in this ice-rich region
could lock up the bulk of this water.

Direct detections of water ice are complicated by the fact that IR
absorption spectroscopy requires a background light source, and thus a
favorable near edge-on orientation of the disk. In addition, care has
to be taken that foreground clouds do not contribute to the water ice
absorption \citep{Pontoppidan05crbr}. The 3 $\mu$m water ice band has
been detected in only a few disks \citep{Terada07,Honda09}. 
To measure the bulk of the ice, one needs
to go to longer wavelengths, where the ice features can be seen in
emission. Indeed, the crystalline H$_2$O features at 45 or 60
$\mu$m have been detected in several sources with ISO-LWS
\citep{Malfait98,Malfait99,Chiang01} and {\it Herschel}-PACS
\citep[][\emph{Bouwman et al.}, in prep]{McClure12}. Quantitatively, the data
are consistent with 25--50\% of the oxygen in water ice on grains in
the emitting layer.

\begin{figure}[t]
\begin{centering}
  \includegraphics[width=0.45\textwidth]{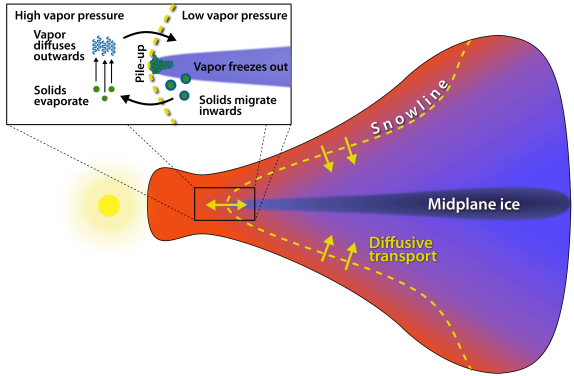}
  \caption{Cartoon illustrating the snow line as a function of radius
    and height in a disk and transport of icy planetesimals across the
    snowline. Diffusion of water vapor from inner to outer disk followed by
    freeze-out results in pile-up of ice just beyond the snowline (the
    cold finger effect).  Figure by M.\ Persson, based on
    \citet{Meijerink09,Ciesla06}. }
  \label{fig:snowline}
\end{centering}
\end{figure}

The ISO-LWS far-infrared spectra also suggested a strong signature of
hydrated silicates in at least one target \citep{Malfait99}. Newer
{\it Herschel}-PACS data show no sign of such a feature in the same
target (\emph{Bouwman}, priv. comm.). An earlier claim of hydrated
silicates at 2.7 $\mu$m in diffuse clouds has now also been refuted
\citep{Whittet98,Whittet10}. Moreover, there is no convincing
detection of any mid-infrared feature of hydrated silicates in
hundreds of {\it Spitzer} spectra of T Tauri
\citep[e.g.,][]{Olofsson09,Watson09}, Herbig Ae
\citep[e.g.,][]{Juhasz10} and warm debris \citep[e.g.,][]{Olofsson12}
disks. Overall, the strong observational consensus is that the
silicates prior to planet formation are `dry'.

\subsection{Chemical models of disks}
\label{sect:diskmodel}

The observations of gaseous water discussed in \S~\ref{sect:diskobs}
indicate the presence of both rotationally hot $T_{\rm ex} \approx 450$~K and
cold ($T_{\rm ex} < 50$~K) water vapor, with abundances of $\sim 10^{-4}$ and much lower values, respectively.  
Based on the chemistry of water vapor discussed in
\S~\ref{sect:chemistry}, we expect it to have a relatively well
understood distribution within the framework of the disk thermal
structure, potentially modified by motions of the various solid or
gaseous reservoirs.  This is broadly consistent with the observations.

Traditionally, the snow line plays a critical role in the distribution
of water, representing the condensation or sublimation front of water
in the disk, where the gas temperatures and pressures allow water to
transition between the solid and gaseous states
(Fig.~\ref{fig:snowline}).  For the solar nebula disk, there is a rich
literature on the topic \citep{Hayashi81, Sasselov04,
  Podolak04, Lecar06, Davis07, Dodson-Robinson09}.  Within our modern
astrophysical understanding, this dividing line in the midplane is
altered when viewed within the framework of the entire disk physical
structure.  There are a number of recent models of the water
distribution that elucidate these key issues \citep{Glassgold09,
  Woitke09, Bethell09, Willacy09,Gorti11, Najita11,
  Vasyunin11,Fogel11, Walsh12,Kamp13}.

\subsubsection{General distribution of gaseous water}

\begin{figure}[t]
\begin{centering}
  \includegraphics[width=0.4\textwidth]{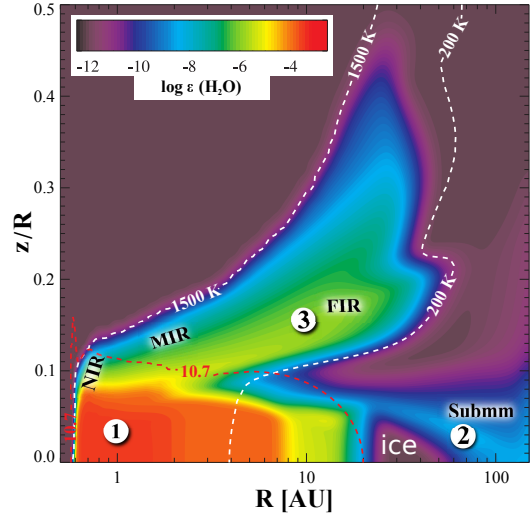}
  \caption{Abundance of gaseous water relative to total hydrogen as a
    function of radial distance, $R$, and relative height above the
    midplane, $z/R$, for a disk around an A-type star ($T_*$=8600 K).
    Three regions with high H$_2$O abundance can be
    distinguished. Regions 1 and 3 involve high-temperature chemistry,
    whereas region 2 lies beyond the snow line and involves
    photodesorption of water ice. The
    white contours indicate gas temperatures of 200 and 1500~K,
    whereas the red contour shows the $n_{\rm H}=5\times 10^{10}$ cm$^{-3}$
    density contour. From \citet{Woitke09}. }
  \label{fig:waterdisk}
\end{centering}
\end{figure}

Fig.~\ref{fig:waterdisk} shows the distribution of water vapor in a
typical kinetic chemical disk model with radius $R$ and height $z$.
The disk gas temperature distribution is crucial for the chemistry.
It is commonly recognized that dust on the disk surface is warmer than
in the midplane due to direct stellar photon heating \citep{Calvet92,
  Chiang97}.  Furthermore the gas temperature is decoupled from the
dust in the upper layers due to direct gas heating
\citep[e.g.,][]{Kamp04}.  There are roughly 3 areas where water vapor
is predicted to be abundant and therefore possibly emissive \citep[see
also discussion in][]{Woitke09h2o}.
These 3 areas or ``regions'' are labelled with coordinates (radial and
vertical) that are specific to the physical structure (radiation
field, temperature, density, dust properties) of this model.
Different models (with similar dust- and gas-rich conditions) find the
same general structure, but not at the exact same physical location.

{\bf Region 1} ($R$ = inner radius to 1.5 AU; $z/R$ $<$ 0.1): this
region coincides with the condensation/sublimation front in the
midplane at the snow line. 
Inside the snow line water vapor will be abundant.  Reaction
timescales imposed by chemical kinetics limit the overall abundance
depending on the gas temperature.  As seen in
Fig.~\ref{fig:waterdisk}, if the gas temperature exceeds $\sim 400$~K
then the midplane water will be quite abundant, carrying all available
oxygen not locked in CO and refractory grains.  If the gas temperature
is below this value, but above the sublimation temperature of $\sim
160$~K, then chemical kinetics could redistribute the oxygen towards
other species.  During the early gas- and dust-rich stages up to a few
Myr, this water vapor dominated region will persist and is seen in
nearly all models.  However, as solids grow, the penetrating power of
UV radiation is increased.  Since water vapor is sensitive to
photodissociation by far-UV, this could lead to gradual
decay of this layer, which would be consistent with the non-detection
of water vapor inside the gaps of a small sample of transition disks
\citep[e.g.,][]{Pontoppidan10,Zhang13}.

{\bf Region 2} ($R >$ 20 AU; surface layers and outer disk midplane):
  In these disk layers the {\em dust} temperature is uniformly below
  the sublimation temperature of water.  Furthermore at these high
  densities ($n > 10^{6}$ cm$^{-3}$) atoms and molecules freeze out on
  dust grains on short timescales (\S 2.4).
Under these circumstances, in the absence of non-thermal desorption
mechanisms, models predict strong freeze-out with the majority of
available oxygen present on grains as water ice.
Much of this may be primordial water ice supplied by the natal cloud
\citep[][Fig.~\ref{fig:diskhistory}]{Visser11}.

The detection of rotationally cold water vapor emission in the outer
disk of TW Hya demonstrates that a tenuous layer of water vapor is
present and that some non-thermal desorption process is active
\citep{Hogerheijde11}.   The
leading candidate is photodesorption of water ice
\citep{Dominik05, Oberg09h2o}, as discussed in \S~\ref{sect:icechemistry},
particularly given the high UV luminosities of T Tauri stars
\citep{Yang12}.  This UV excess is generated by accretion and
dominated by Ly$\alpha$ line emission \citep{Schindhelm12}.

Once desorbed as OH and H$_2$O, the UV radiation then also destroys the
water vapor molecules leading to a balance between these processes and
a peak abundance near $(1-3) \times 10^{-7}$ \citep{Dominik05,
  Hollenbach09}.  In general most models exhibit this layer, which is
strongly dependent on the location and surface area of ice-coated
grains (i.e. less surface area reduces the effectiveness of
photodesorption).  Direct comparison of models with observations finds
that the amount of water vapor predicted to be present exceeds the
observed emission \citep{Bergin10, Hogerheijde11}.  This led to the
suggestion that the process of grain growth and sedimentation could
operate to remove water ice from the UV exposed disk surface layers.
This is consistent with spectroscopic data of the TW Hya scattered
light disk, which do not show water ice features in the spectrum
originating from this layer \citep{Debes13}.  
However, further fine tuning of this settling mechanism is needed 
\citep[\emph{Dominik
  and Dullemond}, in prep.,][]{Akimkin13}. An alternative
explanation may be a smaller dust disk compared with the gas disk
\citep{Qi13}.

{\bf Region 3} ($R <$ 20 AU; $z/R >$ 0.1): Closer to
the exposed disk surface the gas and dust become thermally
decoupled.  The density where this occurs depends on the relative
amount of dust grains in the upper atmosphere, which may be altered by
dust coagulation and settling \citep[][]{Jonkheid04, Nomura07} and on
the thermal accommodation of the dust gas interaction \citep{Burke83}.
In these decoupled layers $T_{\rm gas} \gg T_{\rm dust}$, 
and when the gas temperature exceeds a few hundred K the
neutral-neutral gas-phase pathways for water formation 
become efficient, leading to water abundances of
order $10^{-5}$ (Fig.~\ref{fig:waterdisk}).

More directly, the disk surface is predicted to be water vapor rich at
gas temperatures $\gtrsim$ few hundred K and dust temperatures $\sim
$100~K.  Indeed, there should exist surface layers at radii where
the midplane temperature is sufficiently low to freeze water vapor, but
where the surface can support water formation via
the high-temperature chemistry (i.e., region 3 goes out to
larger radii than region 1).  Thus the water zone on the
disk surface presents the largest surface area and it is this water
that is readily detected with current astronomical observations of
high-lying transitions of H$_2^{16}$O with {\it Spitzer} and {\it
  Herschel}.  The snow line in the midplane is thus potentially hidden
by the forest of water transitions produced by the hot chemistry on
the surface.

\begin{figure}[t]
\begin{centering}
  \includegraphics[width=0.45\textwidth]{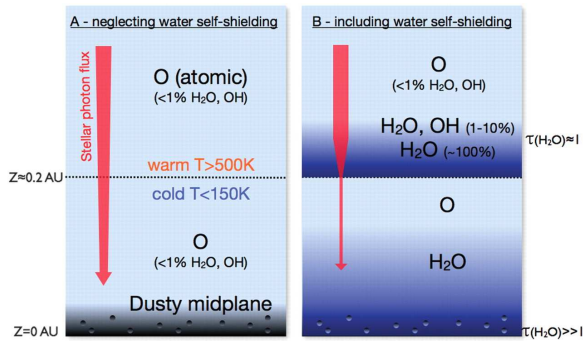}
  \caption{Cartoon illustrating the water self-shielding mechanism and
    the resulting vertical stratification of O and H$_2$O. Inclusion
    of water self-shielding in the upper layers leads to a `wet' warm
    layer. From \citet{Bethell09}. }
  \label{fig:Bergin}
\end{centering}
\end{figure}

There are some key dependences and differences which can be
highlighted.  One important factor is the shape of the UV radiation
field.  In general, models that use a scaled interstellar UV radiation
field, for example based on FUSE/IUE/HST observations of the UV excess
\citep{Yang12}, neglect the fact that some molecules like H$_2$ and CO
require very energetic photons to photodissociate, which are not
provided by very cool stars.  A better approach is to take the actual
stellar continua into account \citep{vanDishoeck06photo}, with UV excess
due to accretion added where appropriate \citep{vanZadelhoff03}.  A
very important factor in this regard is the relative strength of the
Ly$\alpha$ line to the overall UV continuum.  Observations
find that Ly$\alpha$ has nearly an order of magnitude more UV flux
than the stellar FUV continuum in accreting sources \citep{Bergin03,
  Schindhelm12}.  In addition because of the difference in scattering
(Ly$\alpha$ isotropic from H atom surface; UV continuum anisotropic
from dust grains), Ly$\alpha$ will dominate the radiation field deeper
into the disk \citep{Bethell11}.

Most models find the presence of this warm water layer in dust-dominated disks.
However, \citet{Bethell09} suggest that water can form in such high abundances
in the surface layer that it mediates the transport of the energetic
UV radiation by becoming self-shielding (Fig.~\ref{fig:Bergin}).
If this is the case, then the surface water would survive for longer
timescales, because it is somewhat decoupled from the dust evolution.
As a consequence water and chemistry in the midplane might be protected
even as the FUV absorbing dust grains settle to the midplane.

Additional factors of importance for the survival of this emissive surface
layer are the gas temperature and molecular hydrogen abundance.
Although theoretical solutions for the gas temperature are inherently
uncertain, it is clear that hot ($T_{\rm gas} >$ few hundred K) layers
exist on disk surfaces \citep{Bruderer12}.  However, as
the gas disk dissipates, the accretion rate onto the star decays on
timescales of a few Myr.  Thus the UV luminosity that is associated
with this accretion declines and the disk will cool down, cutting off
the production of water from the hot ($T\gtrsim$ 400~K) gas-phase
chemistry on the exposed surface.
 In addition, as shown
by \citet{Glassgold09} and \citet{Adamkovics13}, the formation of
surface water requires the presence of H$_2$ to power the
initiating reaction.  
Finally, vertical mixing through turbulence or disk winds can bring
water ice from the lower to the upper layers where the ice sublimates
and adds to the oxygen budget and water emission \citep{Heinzeller11}.

\subsubsection{Planetesimal formation and water ice transport}

The dust particles in disks collide and grow, with water ice mantles
generally thought to help the coagulation processes. The evolution of
dust to pebbles, rocks and planetesimals (1--100 km bodies, the
precursors of comets and asteroids) is described in the chapters by
{\it Testi et al.} and {\it Johansen et al.}.  The disk models cited
above do not take into account transport of ice-rich planetesimals
from the cold outer to the warm inner disk, even though such radial
drift is known to be highly effective at a few AU for rocks up to
meter size (or mm size further out in the disk)
\citep[e.g.,][]{Weidenschilling93}.  This drift of icy planetesimals
can be a source of water vapor enrichment inside the snow line
\citep{Ciesla06}.  The astrophysical signature of this phenomenon
would be the presence of water vapor in the inner disk with an
abundance greater than the stellar oxygen abundance, because the
planetesimals are hydrogen-poor, that is, the main volatile species,
H$_2$, is not present in water-rich planetesimal ices.  Some of this
hot water can diffuse outwards again and re-condense just outside the
snowline (the cold finger effect. Fig.~\ref{fig:snowline}) which
increases the density of solids by a factor of 2--4 and thereby
assists planet formation \citep{Stevenson88}. Alternatively, icy
grains can be trapped in pressure bumps where they can grow rapidly to
planetesimals before moving inward \citep[e.g.,][]{vanderMarel13}.

\section{\bf WATER IN THE OUTER SOLAR SYSTEM }

In the standard model of the disk out of which our solar system
formed, the snow line was at 2.7 AU at the end of the gas-rich phase
\citep{Hayashi81}. This snow line likely moved inward from larger
distances in the early embedded phase \citep{Kennedy08}. Thus, it is
no surprise that water ice is a major constituent of all solar system
bodies that formed and stayed beyond the snow line.
Nevertheless, their water ice 
content, as measured by the mass in ice with respect to total ice+rock
mass can differ substantially, from $<$1\% for some asteroids to
typically 50\% for comets. Also, the observational signatures of water
on these icy bodies and its isotopic ratio can differ. In the
following sections, we review our knowledge of water ice in the
present-day solar system. In \S~\ref{sect:origin}, possible
mechanisms of supplying water from these reservoirs to the terrestrial
planet zone will be discussed.

\subsection{Outer asteroid belt}

Since the outer asteroid belt is located outside the Hayashi snow
line, it provides a natural reservoir of icy bodies in the solar
system. This part of the belt is dominated by so-called C, P and D
class asteroids with sizes up to a few 100 km at distances of $\sim$3, 4
and $\geq$4 AU, respectivily, characterized
by their particularly red colors and very low albedos, $\lapprox$0.1
\citep{Bus02}. Because of their spectroscopic similarities to the
chemically primitive carbonaceous chondrites found as meteorites on
Earth, C-type asteroids have been regarded as largely unaltered,
volatile-rich bodies. The P- and D-types may be even richer
in organics.

The water content in these objects has been studied though IR
spectroscopy of the 3 $\mu$m band. In today's solar system, 
any water ice on the surface would
rapidly sublimate at the distance of the belt, so only water bonded to the rocky
silicate surface is expected to be detected.  Hydrated minerals can be
formed if the material has been in contact with liquid water.
The majority of the C-type asteroids show hydrated silicate
absorption, indicating that they indeed underwent heating and aqueous
alteration episodes \citep[][and refs.\ therein]{Jones90}. However,
only 10\% of the P and D-type spectra show weak water absorption,
suggesting that they have largely escaped this processing
and that the abundance of hydrated
silicates gradually declines in the outer asteroid belt. Nevertheless,
asteroids that do not display water absorption
on their surfaces (mostly located beyond 3.5 AU) may still retain ices
in their interior. Indeed, water fractions of 5--10\% of their total
mass have been estimated. 
This is consistent with models that show that buried ice can persist
in the asteroid belt within the top few meters of the surface over
billions of years, as long as the mean surface temperature is less
than about 145 K \citep{Schorghofer08}. Water vapor has recently been
detected around the dwarf planet Ceres at 2.7 AU in the asteroid belt, with a
production rate of at least $10^{26}$ mol s$^{-1}$, directly
confirming the presence of water \citep{Kuppers14}.

\citet{Hsieh06} discovered a new population of small objects in the
main asteroid belt, displaying cometary characteristics. These
so-called main belt comets (ten are currently known) display clearly
asteroidal orbits, yet have been observed to eject
dust and thus satisfy the observational definition of a comet. These
objects are unlikely to have originated elsewhere in the solar system
and to have subsequently been trapped in their current
orbits. Instead, they are intrinsically icy bodies, formed and stored
at their current locations, with their cometary activity triggered by
some recent event.

Since main belt comets are optically faint, it is not known whether
they display hydration spectral features that could point to the
presence of water.  Activity of main belt comets is limited to the
release of dust and direct outgassing of volatiles, like for Ceres,
has so far not been detected. The most stringent indirect upper limit
for the water production rate derived from CN observations is that in
the prototypical main belt comet 133P/Elst-Pizarro $<1.3\times
10^{24}$ mol s$^{-1}$ \citep{Licandro11}, which is subject to
uncertainties in the assumed water-to-CN abundance ratio. For
comparison, this is five orders of magnitude lower than the water
production rate of comet Hale-Bopp.  Its mean density is 1.3 gr
cm$^{-3}$ suggesting a moderately high ice fraction \citep{Hsieh04}.
\emph{Herschel} provided the most stringent direct upper limits for
water outgassing in 176P/LINEAR ($<4 \times 10^{25}$ mol s$^{-1}$,
3$\sigma$; \citealt{deValBorro12}) and P/2012 T1 PANSTARRS ($<8 \times
10^{25}$ mol s$^{-1}$; \citealt{ORourke13}).

Another exciting discovery is the direct spectroscopic
detection of water ice on the asteroid 24 Themis
\citep{Campins10,Rivkin10}, the largest (198 km diameter) member of
the Themis dynamical family at $\sim$3.2 AU, which also includes three
main belt comets.  The 3.1 $\mu$m spectral feature detected in Themis
is significantly different from those in other asteroids, meteorites
and all plausible mineral samples available. \citet{Campins10} argue
that the observations can be accurately matched by small ice
particles evenly distributed on the surface. 
 A subsurface ice reservoir could also be present if Themis underwent
differentiation resulting in a rocky core and an ice mantle. 
\citet{Jewitt12b} find no direct evidence of
outgassing from the surface of Themis or Cybele with a 5$\sigma$ upper
limit for the water production rate  $1.3 \times
10^{28}$~mol~s$^{-1}$, assuming a cometary water-to-CN mixing
ratio. 
They conclude that any ice that exists on these bodies should be
relatively clean and confined to a $<$10\% fraction of the
Earth-facing surface.

Altogether, these results suggest that water ice may be common below
asteroidal surfaces and widespread in asteroidal interiors down to
smaller heliocentric distances than previously expected. Their water
contents are clearly much higher than those of meteorites that
originate from the inner asteroid belt, which have only 0.01\% of
their mass in water \citep{Hutson00}.

\subsection{Comets}
\label{sect:comets}

Comets are small solar system bodies with radii less than 20 km that
have formed and remained for most of their lifetimes at large
heliocentric distances. Therefore, they likely contain some of the
least-processed, pristine ices from the solar nebula disk.  They have
often been described as `dirty snowballs', following the model of
\citet{Whipple50}, in which the nucleus is visualized as a conglomerate
of ices, such as water, ammonia, methane, carbon dioxide, and carbon
monoxide, combined with meteoritic materials. However, \emph{Rosetta}
observations during the Deep Impact encounter \citep{Kuppers05}
suggest a dust-to-gas ratio in excess of unity in comet 9P/Tempel 1.
Typically, cometary ice/rock ratios are of order unity
with an implied porosity well over 50\% \citep{AHearn11}.

The presence and amount of water in comets is usually quantified
through their water production rates, which are traditionally inferred
from radio observations of its photodissociation product, OH, at 18~cm
\citep{Crovisier02}. Measured rates vary from $10^{26}$ to $10^{29}$
mol s$^{-1}$. The first direct detection of gaseous water in comet
1P/Halley, through its $\nu_3$ vibrational band at 2.65 $\mu$m, was
obtained using the KAO \citep{Mumma86}. The 557 GHz transition of
ortho-water was observed by \emph{SWAS} \citep{Neufeld00comet} and
$Odin$ \citep{Lecacheux03}, whereas \emph{Herschel} provided for the
first time access to multiple rotational transitions of both ortho-
and para-water \citep{Hartogh11}. These multi-transition mapping
observations show that the derived water production rates are
sensitive to the details of the excitation model used, in particular
the ill-constrained temperature profile within the coma, with
uncertainties up to 50\% \citep{Bockelee12}.

Dynamically, comets can be separated into two general groups:
short-period, Jupiter-family comets and long-period comets (but see
\citealt{Horner03} for a more detailed classification). Short-period
comets are thought to originate from the Kuiper belt, or the
associated scattered disk, beyond the orbit of Neptune, while
long-period comets formed in the Jupiter-Neptune region and were
subsequently ejected into the Oort cloud by gravitational interactions
with the giant planets. In reality, the picture is significantly more
complex due to migration of the giant planets in the early solar
system (see below).
In addition, recent simulations \citep{Levison10} suggest that the Sun may have
captured comets from other stars in its birth cluster. In this case, a
substantial fraction of the Oort-cloud comets, perhaps in excess of
90\%, may not even have formed in the Sun's protoplanetary
disk. Consequently, there is increasing emphasis on classifying comets
based on their chemical and isotopic composition rather than orbital
dynamics \citep{Mumma11}. There is even evidence for heterogeneity
within a single comet, illustrating that comets may be built up from
cometesimals originating at different locations in the disk
\citep{AHearn11}.

Traditionally, the ortho-to-para ratio in water and other cometary
volatiles has been used to contrain the formation and thermal history
of the ices.  Recent laboratory experiments suggest that this ratio is
modified by the desorption processes, both thermal sublimation and
photodesorption, and may therefore tell astronomers less about the
water formation location than previously thought (see discussions in
\citealt{vanDishoeck13} and \citealt{Tielens13}).

\subsection{Water in the outer satellites}

Water is a significant or major component of almost all 
moons of the giant planets for which densities or spectral
information are available. Jupiter's Galilean moons exhibit
a strong gradient from the innermost (Io, essentially all rock, no ice
detected) to the outermost (Callisto, an equal mixture of rock and
ice). Ganymede has nearly the same composition and hence rock-to-ice
ratio as Callisto. Assuming that the outermost of the moons reflects
the coldest part of the circumplanetary disk out of which the moons
formed, and hence full condensation of water, Callisto's (uncompressed)
density matches that of solar-composition material in which the
dominant carbon-carrier was methane rather than carbon monoxide
\citep{Wong08}.  
The fact that Callisto has close to
(but not quite) the full complement of water expected based on the
solar oxygen abundance implies that the disk around Jupiter had a
different chemical composition (CH$_4$-rich, CO-poor) from that of the
solar nebula disk, which was CO-dominated \citep{Prinn81}.

The Saturnian satellites are very different. For satellites large
enough to be unaffected by porosity, but excluding massive Titan, the
ice-to-rock ratio is higher than for the Galilean moons
\citep{Johnson09}.  However, Titan---by far the most massive moon---has 
a bulk density and mass just in between, and closely resembling,
Ganymede's and Callisto's. Evidently the Saturnian satellite system
had a complex collisional history, in which the original ice-rock ratio
of the system was not preserved except perhaps in Titan.
The moon Enceladus, which exhibits volcanic and geyser activity,
offers the unique opportunity to sample Saturnian system water
directly.  Neptune's Triton, like Pluto, has a bulk density and hence
ice-rock ratio consistent with what is expected for a solar nebula
disk in which CO dominated over CH$_4$.  Its water fraction is about
15--35\%.  At this large distance from the Sun, N$_2$ can also be
frozen out and Triton's spectrum is indeed dominated by N$_2$ ice with
traces of CH$_4$ and CO ices; the water signatures are much weaker
than on other satellites \citep{Cruikshank00}.
The smaller Trans Neptunion Objects (TNOs) ($<$few hundred km size) are
usually found to have mean densities
around 1 g cm$^{-3}$ and thus a high ice fraction.  Larger TNOs such
as Quaoar and Haumea have much higher densities (2.6--3.3
g\,cm$^{-3}$) suggesting a much lower ice content, even compared with
Pluto (2.0 g\,cm$^{-3}$) \citep{Fornasier13}.

In summary, the water ice content of the outer satellites varies with
position and temperature, not only as a function of distance from the
Sun, but also from its parent planet. Ice fractions are
generally high ($\geq$50\%) in the colder parts and consistent with solar
abundances depending on the amount of oxygen locked up in CO. However,
collisions can have caused a strong reduction of the water ice
content.

\subsection{Water in the giant planets}

The largest reservoirs of what was once water ice in the solar nebula
disk are presumably locked up in the giant planets. 
If the formation of giant planets started 
with an initial solid core
of 10--15 Earth masses
with subsequent growth from a swarm of planetesimals of ice
and rock with solar composition, 
the giant planets could have had several Earth masses of oxygen, some
or much of which may have been in water molecules in the original
protoplanetary disk. The core also gravitationally attracts
the surrounding gas in the disk consisting mostly of H and He with the
other elements in solar composition. The resulting gas giant
planet has a large mass and diameter, but a low overall density
compared with rocky planets. The giant planet atmosphere is
expected to have an excess in heavy elements, either due
to the vaporization of the icy planetesimals when they entered the
envelopes of the growing planet during the heating phase, or due to
partial erosion of the original core, or both 
\citep{Encrenaz08,Mousis09}.  These calculations assume that all heavy
elements are equally trapped within the ices initially, which is a
debatable assumption, and that the ices fully evaporate, with most of
the refractory material sedimenting onto the core.  In principle, the
excesses provide insight into giant planet formation mechanisms and
constraints on the composition of their building blocks.

\subsubsection{Jupiter and Saturn}

For Jupiter, elemental abundances can be derived from spectroscopic
observations and from data collected by the Galileo probe, which
descended into the Jovian atmosphere. Elements like C, N, S and the
noble gases Ar, Kr and Xe, have measured excesses as expected at
4$\pm$2 \citep{Owen06}. However, O appears to show significant
depletion.  Unfortunately, the deep oxygen abundance in Jupiter is not
known. The measured abundance of gaseous water---the primary carrier
of oxygen in the Jovian atmosphere since there is little CO---provides
only a lower limit since the troposphere at about 100 mbar is a region
of minimum temperature ($\sim$110 K for Jupiter) and therefore acts as
a cold trap where water can freeze out. Thus, the amount of gaseous
water is strongly affected by condensation and rainout associated with
large-scale advective motions \citep{Showman00}, meteorological
processes \citep{Lunine87}, or both.  The Galileo probe fell into a
so-called `hot spot' (for the excess brightness observed in such
regions at 5 $\mu$m wavelengths), with enhanced transparency and hence
depleted in water, and is thus not representative of the planet as a
whole.  The water abundance, less than 1/10 the solar value in the
upper atmosphere, was observed to be higher at higher pressures,
toward the end of the descent \citep{Roos04}. The sparseness of the
measurements made it impossible to know whether the water had `leveled
out' at a value corresponding to 1/3 solar or would have increased
further had the probe returned data below the final 21 bar level.

As noted above, predictions for standard models of planetesimal
accretion---where volatiles are either
adsorbed on, or enclathrated in, water ice---give oxygen abundances
3--10 times solar in the Jovian deep interior.  Although atmospheric
explanations for the depleted water abundance in Jupiter are
attractive, one must not rule out the possibility that water truly is
depleted in the Jovian interior---that is, the oxygen-to-hydrogen ratio
in Jupiter is less enriched than the carbon value at 4$\pm$2
times solar. A motivation for making such a case is that at least one
planet with a C/O$>$1---a `carbon-rich planet'---has been discovered
\citep{Madhusudhan11a}, companion to the star WASP12a with a C/O
ratio of 0.44, roughly solar. One explanation is that the portion of
this system's protoplanetary disk was somehow depleted in water at the
time the planet formed and acquired its heavy element inventory
\citep{Madhusudhan11b}.

Prior to this discovery, the possibility of a carbon-rich Jupiter was
considered on the basis of the Galileo results alone by
\citet{Lodders04} who proposed that in the early solar system
formation the snow line might have been further from the Sun than
the point at which Jupiter formed, and volatiles adhering to solid
organics rather than water ice were carried into Jupiter.

\citet{Mousis12} looked at the possibility that Jupiter may have
acquired planetesimals from an oxygen-depleted region by examining
element-by-element the fit to the Galileo probe data of two
contrasting models: one in which the planetesimal building blocks of
Jupiter derived from a disk with C/O= 1/2 (roughly, the solar value),
and the other in which C/O=1. Within the curent error bars, 
the two cases cannot be distinguished. However, any
determination of the deep oxygen abundance in Jupiter yielding an
enrichment of less than 2 times solar would be a strong argument in
favor of a water depletion, and thus high C/O ratio, at certain places
and times in the solar nebula disk.

Are such depletions plausible? A wide range of oxidation states
existed in the solar nebula disk at different times and locations. For
example, the driest rocks, the enstatite chondrites, are thought to
have come from parent bodies formed inward of all the other parent
bodies, and their mineralogy suggests reducing conditions---consistent
with a depleted water vapor abundance---in the region of the nebula
where they formed \citep{Krot00}. 
One recent model of the early evolution of Jupiter and Saturn hypothesizes
that these giant planets moved inward significantly during the late
stages of their formation, reaching 1.5 AU in the case of Jupiter
\citep{Walsh11}, almost certainly inward of the snow line (see
\S~\ref{sect:delivery}). If planetesimals in this region accreted
volatiles on refractory organic and silicate surfaces which were then
incorporated into Jupiter, the latter would appear carbon-rich and
oxygen-depleted. However, temperatures in that region may not have
been low enough to provide sufficient amounts of the more volatile
phases. 
A second possibility is that
Jupiter's migration scattered these water poor planetesimals to the
colder outer solar system, where they trapped noble gases and
carbon-and nitrogen-bearing species at lower temperature---but water
ice, already frozen out, was not available.

In order to test such models, one must directly determine the
abundance of oxygen-bearing species in Jupiter and if possible, in
Saturn. The case of Saturn is similar to that of Jupiter in the sense
that the carbon excess as derived from CH$_4$ spectroscopy is as
expected, but again no reliable oxygen abundance can be
determined. Under the conditions present in the Jovian envelope at
least (if not its core), water will dominate regardless of the initial
carbon oxidation state in the solar nebula. NASA's {\it Juno} mission
to Jupiter will measure the water abundance down to many tens of bars
via a microwave radiometer (MWR) \citep{Janssen05}.  Complementary to
the MWR is a near-infrared spectrometer JIRAM (Jovian Infrared Auroral
Mapper), that will obtain the water abundance in the meteorological
layer \citep{Adriani08}. The two instruments together will be able to
provide a definitive answer for whether the water abundance is below
or above solar, and in the latter case, by how much, when {\it Juno}
arrives in 2016. {\it Juno} will also determine the mass of the heavy
element core of Jupiter, allowing for an interpretation of the
significance of the envelope water abundance in terms of total oxygen
inventory.

There is no approved mission yet to send a probe into 
Saturn akin to Galileo. 
However, the {\it Cassini} Saturn
Orbiter will make very close flybys of Saturn starting in 2016,
similar to what {\it Juno} will do at Jupiter.
Unfortunately, a microwave radiometer akin to that on {\it Juno} is
not present on {\it Cassini}, but a determination of the heavy element
core mass of Saturn may be obtained from remote sensing.
In summary, the fascinating possibility that Jupiter and Saturn may
have distinct oxygen abundances because they sampled at different
times and to differing extents regions of the solar nebula disk 
heterogeneous in
oxygen (i.e., water) abundance is testable if 
the bulk oxygen abundances can be measured.

\subsubsection{Uranus and Neptune}

The carbon excesses measured from CH$_4$ near-infrared spectroscopy
give values of 30--50 for Uranus and Neptune, compared with 9 and 4
for Saturn and Jupiter \citep{Encrenaz08}. These values are consistent
with the assumption of an initial core of 10--15 Earth masses with
heavy elements in solar abundances.  Unfortunately, nothing is known
about the water or oxygen content in Uranus and Neptune, since water
condensation occurs at such deep levels that even tropospheric water
vapor cannot be detected.  Unexpectedly, ISO detected water vapor in
the upper atmospheres of both ice giants, with mixing ratios orders of
magnitude higher than the saturation level at the temperature
inversion \citep{Feuchtgruber97}. These observations can only be
accounted for by an external flux of water molecules, due to
interplanetary dust, or sputtering from rings or satellites.
Similarly, {\it Herschel} maps of water of Jupiter demonstrate that
even 15 years after the Shoemaker-Levy 9 impact, 
more than 95\% of the stratospheric
Jovian water comes from the impact
\citep{Cavalie13}. On the other hand, the low D/H ratios in molecular
hydrogen measured by {\it Herschel} in the atmospheres of Uranus and
Neptune 
may imply a lower ice mass fraction of their cores than previously
thought (14--32\%) \citep{Feuchtgruber13} .

It is unlikely that information on the deep oxygen abundance will be
available for Uranus and Neptune anytime soon, since reaching below
the upper layers to determine the bulk water abundance is very
difficult. Thus, data complementary to that for Jupiter and Saturn
will likely come first from observations of Neptune-like exoplanets.

\begin{figure}[t]
  \includegraphics[width=0.45\textwidth]{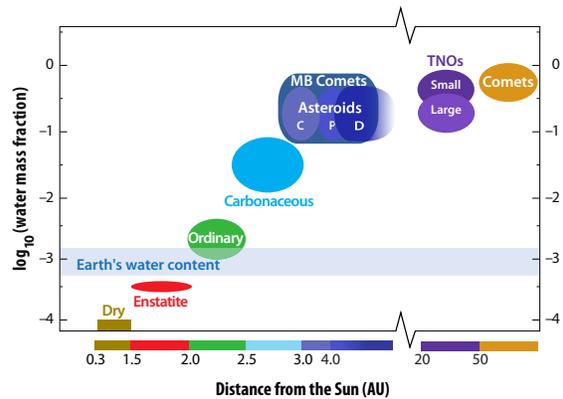}
  \caption{Water content of various parent bodies at the time of
    Earth's growth as a function of radial distance from the
    Sun. Enstatite chondrites originating from asteroids around 1.8 AU
    in the inner disk are very dry.
    In contrast, carbonaceous
    chondrites originating from the outer asteroid belt and beyond
    have a water content of 5--10\%, and main belt comets, TNOs and 
    regular comets even
    more. Figure by M.\ Persson, adapted from \citet{Morbidelli12}. }
  \label{fig:asteroids}
\end{figure}

\section{\bf WATER IN THE INNER SOLAR SYSTEM}

Terrestrial planets were formed by accretion of rocky planetesimals,
and they have atmospheres that are only a small fraction of their
total masses. They are small and have high densities compared with
giant planets. Earth and Venus have very similar sizes and masses,
whereas Mars has a mass of only 10\% of that of Earth and a somewhat
lower density (3.9 vs 5.2 gr cm$^{-3}$).
Terrestrial planets undergo different evolutionary processes compared
with their gas-rich counterparts. In particular, their atmospheres
result mostly from outgassing and from external bombardment.  Internal
differentiation of the solid material after formation leads to a
structure in which the heavier elements sink to the center, resulting
in a molten core (consisting of metals like iron and nickel), a mantle
(consisting of a viscous hot dense layer of magnesium rich silicates)
and a thin upper crust (consisting of colder rocks).

Water has very different appearances on Venus, Earth and Mars, which
is directly related to their distances from the Sun at 0.7,
1.0 and 1.7 AU, respectively. On Venus, with surface temperatures
around 730 K, only gaseous water is found, whereas the surface of Mars
has seasonal variations ranging from 150--300 K resulting in water
freezing and sublimation. Its current mean surface pressure of 6 mbar
is too low for liquid water to exist, but there is ample evidence for
liquid water on Mars in its early history. Most of the water currently
on Mars is likely subsurface in the crust down to 2 km depth. Earth is
unique in that its mean surface temperature of 288 K and pressure of 1
bar allow all three forms of water to be present: vapor, liquid and
ice. In \S~\ref{sect:origin}, the origin of water on these three planets
will be further discussed. It is important to keep in mind that even
though these planets have very different atmospheres today, they
may well have started out with comparable initial water mass fractions
and similar atmospheric compositions dominated by H$_2$O, CO$_2$, and
N$_2$ \citep{Encrenaz08}. 

Earth has a current water content that
is non-negligible.
The mass of the water contained in the Earth's crust (including the
oceans and the atmosphere) is $2.8 \times 10^{-4}$ Earth masses,
denoted as `one Earth ocean' because almost all of this is in the
surface waters of the Earth. The mass of the water in the present-day
mantle is uncertain. \citet{Lecuyer98} estimate it to be in the range
of (0.8--8)$\times 10^{-4}$ Earth masses, equivalent to 0.3--3 Earth
oceans.  More recently, \citet{Marty12} provides arguments in favor of
a mantle water content as high as $\sim$7 Earth oceans. However, an
even larger quantity of water may have resided in the primitive Earth
and been subsequently lost during differentiation and impacts. Thus,
the current Earth has a water content of roughly 0.1\% by mass, larger than
that of enstatite chondrites, and it is possible that the primitive
Earth had a water content comparable with or larger than that of
ordinary chondrites, definitely larger than the water content of
meteorites and material that condensed at 1~AU
(Fig.~\ref{fig:asteroids}).

Remote sensing and space probes have detected water in the lower
atmosphere of Venus at a level of a few tens of parts per millions
\citep{Gurwell07}, confirming that it is a minor component. Radar
images show that the surface of Venus is covered with volcanoes and is
likely very young, only 1 billion years.  Direct sampling of the
Venusian crust is needed to determine the extent of past and present
hydration which will provide an indication of whether Venus once had
an amount of water comparable to that on Earth. Hints that there may
have been much more water on Venus and Mars come from D/H measurements
(see \S~\ref{sect:hdo}).

\section{\bf ORIGIN OF WATER TERRESTRIAL PLANETS} 
\label{sect:origin}

Earth, Mars and perhaps also Venus all have water mass fractions that
are higher than those found in meteorites in the inner solar nebula
disk.  Where did this water come from, if the local planetesimals were
dry? There are two lines of arguments that provide clues to its
origin.  One clue comes from measuring the D/H ratio of the various
water mass reservoirs.  The second argument looks at the water mass
fractions of various types of asteroids and comets, combined with the
mass delivery rates of these objects on the young planets at the time
of their formation, based on models of the dynamics of these
planetesimals. For example, comets have plenty of water, but the
number of comets that enter the inner 2 AU and collide with
proto-Earth or -Mars is small. To get one Earth ocean, $\sim 10^8$
impacting comets are needed. It is therefore not a priori obvious
which of the reservoirs shown in Fig.~\ref{fig:asteroids}
dominates. 

Related to this point is the question whether the planets formed `wet'
or `dry'. In the wet scenario, the planets either accreted a
water-rich atmosphere or they formed from planetesimals with water
bonded to silicate grains at 1 AU, with subsequent outgassing of water
as the material is heated up during planet formation. In the dry
scenario, the terrestrial planets are initially built up from
planetesimals with low water mass fractions and water is delivered to
their surfaces by water-rich planetesimals, either late in the
building phase, or even later after the planets have formed and
differentiated. If after differentiation, this scenario is often
called a `late veneer'.  In the following, we first discuss the D/H
ratios and then come back to the mass fraction arguments.

\subsection{D/H ratios in solar system water reservoirs}
\label{sect:hdo}

\begin{figure}[t]
\begin{centering}
\includegraphics[angle=0,width=0.45\textwidth]{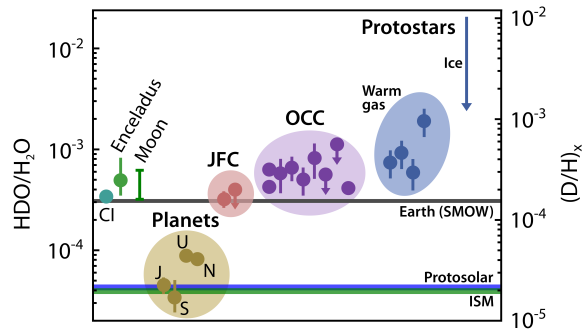}
\caption{D/H ratio in water in comets and warm protostellar envelopes
  compared to values in the Earth oceans, the giant planets, the solar
  nebula disk, and the interstellar medium. Values for carbonaceous
  meteorites (CI), the Moon and Saturn's moon Enceladus are presented
  as well. Note that both (D/H)$_{\rm X}$, the deuterium to hydrogen
  ratio in molecule X, and HDO/H$_2$O are plotted; measured HD/H$_2$
  and HDO/H$_2$O ratios are 2$\times$(D/H)$_{\rm X}$.  Figure by
  M. Persson, based on \citet{Bockelee12} and subsequent measurements
  cited in the text. The protostellar data refer to warm gas
  \citep[][and subm.]{Persson13}}
\label{fig:dh}
\end{centering}
\end{figure}

The D/H ratio of water potentially provides a unique fingerprint of
the origin and thermal history of water. Figure~\ref{fig:dh}
summarizes the various measurements of solar system bodies, as well as
interstellar ices and protostellar objects.

The primordial [D]/[H] ratio set shortly after the Big Bang is
$2.7\times 10^{-5}$.  Since then, deuterium has been lost due to
nuclear fusion in stars, so the deuterium abundance at the time of our
solar system formation, 4.6 billion years ago (redshift of about
$z\approx 1.4$), should be lower than the primordial value, but higher
than the current day interstellar [D]/[H] ratio. The latter value has
been measured in the diffuse local interstellar medium from UV
absorption lines of atomic D and H, and varies from place to place but
can be as high as $2.3\times 10^{-5}$ \citep{Prodanovic10}. These
higher values suggest that relatively little deuterium has been
converted in stars, so a [D]/[H] value of the solar nebula disk only
slightly above the current ISM value is expected at the time of solar
system formation.  The solar nebula [D]/[H] ratio can be measured from the
solar wind composition as well as from HD/H$_2$ in the atmospheres of
Jupiter and Saturn and is found to be $(2.5\pm0.5)\times 10^{-5}$
\citep[][and references cited]{Robert00}, indeed 
in-between the primordial and the current ISM ratios.\footnote{Note
  that measured HD/H$_2$ and HDO/H$_2$O ratios are 2$\times$(D/H)$_{\rm
    X}$, the D/H ratios in these molecules.}

The D/H ratio of Earth's ocean water is $1.5576 \times 10^{-4}$
(`Vienna Standard Mean Ocean Water', VSMOW or SMOW).
Whether this value is representative of the bulk of Earth's water
remains unclear, as no measurements exist for the mantle or the core.
It is thought that recycling of water in the deep mantle does not
significantly change the D/H ratio.   In any
case, the water D/H ratio is at least a
factor of 6 higher than that of the gas out of which our solar system
formed. Thus, Earth's water must have undergone fractionation
processes that enhance deuterium relative to hydrogen at
some stages during its history, of the kind described in
\S~\ref{sect:chemistry} for interstellar chemistry at low
temperatures.

Which solar system bodies show D/H ratios in water similar to those
found in Earth's oceans? The highest [D]/[H] ratios are found in two
types of primitive meteorites \citep{Robert00}: LL3, an ordinary
chondrite that may have the near-Earth asteroid 433 Eros as its parent
body (up to a factor of 44 enhancement compared to the protosolar
ratio in H$_2$) and some carbonaceous chondrites (a factor of 15 to 25
enhancement). Most chondrites show lower enhancements than
the most primitive meteorites.  Pre-\emph{Herschel} observations of
six Oort-cloud or long-period comets give a D/H ratio in
water of $\sim 3\times 10^{-4}$, a factor of 12 higher than the
protosolar ratio in H$_2$ \citep{Mumma11}.

\emph{Herschel} provided the first measurement of the D/H ratio in a
Jupiter-family comet. A low value of $(1.61\pm 0.24) \times 10^{-4}$,
consistent with VSMOW, was measured in comet 103P/Hartley~2
\citep{Hartogh11}. In additon, a relatively low ratio of $(2.06 \pm
0.22) \times 10^{-4}$ was found in the Oort-cloud comet C/2009 P1
Garradd \citep{Bockelee12} and a sensitive upper limit of $<2 \times
10^{-4}$ (3$\sigma$) was obtained in another Jupiter-family comet
45P/Honda-Mrkos-Pajdu\v{s}\'{a}kov\'{a} (Lis et al.\ 2013).  The
Jupiter-family comets 103P and 45P are thought to originate from the
large reservoir of water-rich material in the Kuiper belt or scattered
disk at 30--50 AU. In contrast, Oort cloud comets, currently at much
larger distances from the Sun, may have formed closer in, near the
current orbits of the giant planets at 5--20 AU, although this
traditional view has been challened in recent years (see
\S~\ref{sect:comets}).  Another caveat is that isotopic ratios in
cometary water may have been altered by the outgassing process
\citep{Brown11}.  Either way, the \emph{Herschel} observations
demonstrate that the earlier high D/H values are not representative of
all comets.

How do these solar system values compare with interstellar and
protostellar D/H ratios? Measured D/H ratios in water in protostellar
envelopes vary strongly. In the cold outer envelope, D/H values for
gas-phase water are as high as $10^{-2}$
\citep{Liu11,Coutens12,Coutens13}, larger than the upper limits in
ices of $<$(2--5)$\times 10^{-3}$ obtained from infrared spectroscopy
\citep{Dartois03,Parise03}. In the inner warm envelope, previous
discrepancies for gaseous water appear to have been resolved in favor of the
lower values, down to (3--5)$\times 10^{-4}$ \citep[][and
subm.]{Jorgensen10hdo,Visser13,Persson13} The latter values are within
a factor of two of the cometary values.

Based on these data, the jury is still out whether the D/H ratio in
solar system water was already set by ices in the early (pre-)collapse
phase and transported largely unaltered to the comet-forming zone
(cf.\ Fig.~\ref{fig:cartoon}), or whether further alteration of D/H
took place in the solar nebula disk along the lines described in
\S~\ref{sect:deuteration} (see also chapter by {\it Ceccarelli et
  al.}).
The original D/H ratio in ices may even have been reset early in the
embedded phase by thermal cycling of material due to accretion events
onto the star (see chapter by {\it Audard et al.}).  Is the high value
of $10^{-2}$ found in cold gas preserved in the material entering the
disk or are the lower values of $<10^{-3}$ found in hot cores and ices
more representative? Or do the different values reflect different
D/H ratios in layered ices \citep{Taquet13}?  

Regardless of the precise initial value, models have shown that
vertical and radial mixing within the solar nebula disk reduces the
D/H ratios from initial values as high as $10^{-2}$ to values as low
as $10^{-4}$ in the comet-forming zones
\citep{Willacy09,yca12,Jacquet13,Furuya13,Albertsson14}.
For water on Earth, the fact that the D enrichment is a factor
$\sim$6 above the solar nebula value rules out the warm thermal
exchange reaction of H$_2$O + HD $\to$ HDO + H$_2$ in the inner nebula
($\sim$1 AU) as the sole cause. Mixing with some cold reservoir with
enhanced D/H at larger distances is needed.

Figure~\ref{fig:dh} contains values for several other solar system
targets. 
Enceladus, one of Saturn's moons with volcanic activity, has a high D/H ratio
consistent with it being built up from outer solar system
planetesimals. Some measurements indicate that lunar water may have a
factor of two higher hydrogen isotopic ratio than the Earth's oceans
\citep{Greenwood11}, although this has been refuted \citep{Saal13}.

Mars and Venus have interestingly high D/H ratios. The D/H ratio of
water measured in the Martian atmosphere is 5.5 times VSMOW,
which is interpreted to imply a significant loss of water over Martian
history and associated enhancement of deuterated water.  Because D is
heavier than H, it escapes more slowly, therefore over time the
atmosphere is enriched in deuterated species like HDO. A
time-dependent model for the enrichment of deuterium on Mars 
assuming a rough outgassing efficiency of 50\%
suggests that the amount of water that Mars must have accreted is
0.04--0.4 oceans. The amount of outgassing may be tested by Curiosity
Mars rover measurements of other isotopic ratios, such as $^{38}$Ar to
$^{36}$Ar.

The extremely high D/H ratio for water measured in the
atmosphere of Venus, about 100 times VSMOW, is almost certainly a
result of early loss of substantial amounts of crustal waters,
regardless of whether the starting value was equal to VSMOW or twice
that value. The amount of water lost is very uncertain because the
mechanism and rate of loss affects the deuterium fractionation
\citep{Donahue92}. Solar wind stripping, hydrodynamic escape,
and thermal (Jeans') escape have very different efficiencies for the
enrichment of deuterium per unit amount of water lost. Values ranging
from as little as 0.1\% to one Earth ocean have been proposed.

\subsection{Water delivery to the terrestrial planet zone}
\label{sect:delivery}

\subsubsection{Dry scenario}

The above summary indicates at least two reservoirs of water ice rich
material in the present-day solar system with D/H ratios consistent
with that in Earth's oceans: the outer asteroid belt and the
Kuiper belt or scattered disk (as traced by comets 103P and 45P).
As illustrated in Fig.~\ref{fig:asteroids},
these same reservoirs also have a
high enough water mass fraction to deliver the overall water content
of terrestrial planets.  The key question is then whether the delivery
of water from these reservoirs is consistent with the current
understanding of the early solar system dynamics.

Over the last decade, there has been increasing evidence that the
giant planets did not form and stay at their current location in the
solar system but migrated through the protosolar disk. The Grand Tack
scenario \citep{Walsh11} invokes movement of Jupiter just after it
formed in the early, gas-rich stage of the disk (few Myr). The Nice
model describes the dynamics and migration of the giant planets in the
much later, gas-poor phase of the disk, some 800--900 Myr after
formation \citep{Gomes05,Morbidelli05,Tsiganis05}, see chapter by
{\it Raymond et al}.

\begin{figure}[t]
\begin{centering}
\includegraphics[angle=0,width=0.45\textwidth]{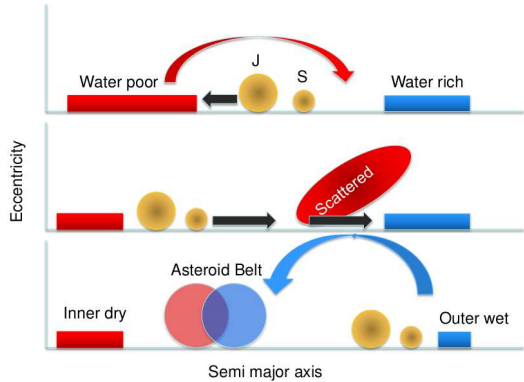}
\caption{Cartoon of the delivery of water-rich planetesimals in the
  outer asteroid belt and terrestrial planet-forming zone based on the
  Grand Tack scenario in which Jupiter and Saturn first move inwards
  to 1.5 AU and then back out again. Figure based on
  \citet{Walsh11}.}
\label{fig:mixing}
\end{centering}
\end{figure}

The Grand Tack model posits that if Jupiter formed earlier than Saturn
just outside the snow line, both planets would have first migrated
inward until they got locked in the 3:2 resonance, then they would
have migrated outward together until the complete disappearance of the
gas disk (see cartoon in Fig.~\ref{fig:mixing}). As summarized by
\citet{Morbidelli12}, the reversal of Jupiter's migration at 1.5~AU
provides a natural explanation for the outer edge of the inner disk of
(dry) embryos and planetesimals at 1 AU, which is required to explain
the low mass of Mars. The asteroid belt at 2--4 AU is first fully
depleted and then repopulated by a small fraction of the planetesimals
scattered by the giant planets during their formation. The outer part
of this belt around 4 AU would have been mainly populated by planetesimals
originally beyond the orbits of the giant planets ($>$5 AU), which
explains similarities between primitive asteroids (C-type) and
comets. Simulations by \citet{Walsh11} show that as the outer asteroid
belt is repopulated, $(3-11)\times 10^{-2}$ Earth masses of C-type
material enters the terrestrial planet region, which exceeds by a
factor 6--22 the minimum mass required to bring the current amount of
water to the Earth. In this picture, Kuiper belt objects would at best
be only minor contributors to the Earth's water budget.

The Nice model identifies Jupiter and Saturn crossing their 1:2
orbital resonance as the next key event in the dynamical evolution of
the disk in the gas-poor phase. After the resonance crossing time at
$\sim 880$ Myr, the orbits of the ice giants, Uranus and Neptune,
became unstable. They then disrupted the outer disk and scattered
objects throughout the solar system, including into the terrestrial
planet region. However, \citet{Gomes05} estimate the amount of
cometary material delivered to the Earth to be only about 6\% of the
current ocean mass. A larger influx of material from the asteroid belt
is expected, as resonances between the orbits of asteroids and giant
planets can drive objects onto orbits with eccentricities and
inclinations large enough to allow them to evolve into the inner solar
system. By this time, the terrestrial planets should have formed
already, so this material would be part of a `late delivery'.

The Grand Tack model provides an attractive, although not
unchallenged, explanation for the observed morphology of the inner
solar system and the delivery of water to the Earth. In the asteroidal
scenario, water is accreted during the formation phase of the
terrestrial planets and not afterwards through bombardment as a late
veneer. Typically, 50\% of the water is accreted after the Earth has
reached 60--70\% of its final mass \citep{Morbidelli12}. This appears
in contradiction with the measurements of the distinct D/H ratio in
lunar water mentioned above, approximately twice that in the Earth's
oceans \citep{Greenwood11}. If valid, this would indicate that a
significant delivery of cometary water to the Earth-Moon system
occurred shortly after the Moon-forming impact. The Earth water would
thus be a late addition, resulting from only one, or at most a few
collisions with the Earth that missed the Moon \citep{Robert11}.
Expanding the sample of objects with accurate D/H measurements is thus
a high priority, long-term science goal for the new submillimeter
facilities.

While the case for late delivery of water on Earth is still open,
\citet{Lunine03} showed that the abundance of water derived for early
Mars is consistent with the general picture of late delivery by comets
and bodies from the early asteroid belt. Unlike Earth, however, the
small size of Mars dictates that it acquired its water primarily from
very small bodies---asteroidal (sizes up to 100 km)---rather than
lunar ($\sim 1700$ km) or larger. The stochastic nature of the
accretion process allows this to be one outcome out of many, but the
Jupiter Grand Tack scenario provides a specific mechanism for removing
larger bodies from the region where Mars formed, thereby resulting in
its small size.

\subsubsection{Wet scenario}

There are two ways for Earth to get its water locally around 1 AU
rather than through delivery from the outer solar system. The first
option is that local planetesimals have retained some water at high
temperatures through chemisorption onto silicate grains
\citep{Drake05,deLeeuw10}.  There is no evidence for such material
today in meteorites nor in the past in interstellar and protoplanetary
dust (\S~\ref{sect:diskobs}). However, the solid grains were bathed in
abundant water vapor during the entire lifetime of gas-rich
disks. This mechanism should differentiate between Earth and
Venus. Computation and lab studies \citep{Stimpfl06} suggest that
chemisorption may be just marginally able to supply the inventory
needed to explain Earth's crustal water, with some mantle water, but
moving inward to 0.7 AU the efficiency of chemisorption should
be significantly lower. Finding that Venus was significantly dryer than
the Earth early in the solar system history would argue for the
local wet source model, although such a model may also overpredict the
amount of water acquired by Mars.

The second wet scenario is that Earth accretes a water rich atmosphere
directly from the gas in the inner disk.  This would require Earth to
have had in the past a massive hydrogen atmosphere (with a molar ratio
H$_2$/H$_2$O larger than 1) that experienced a slow hydrodynamical
escape \citep{Ikoma06}.  One problem with this scenario is that
the timescales for terrestrial planet formation are much longer than
the lifetime of the gas disk. Also, the D/H ratio would be too low,
unless photochemical processes at the disk surface enhance D/H in
water and mix it down to the midplane \citep{Thi10} or over long timescales
via mass-dependent atmospheric escape \citep{Genda08}.

\section{\bf EXOPLANETARY ATMOSPHERES}

H$_2$O is expected to be one of the dominant components in the
atmospheres of giant exoplanets, so searches for signatures of water
vapor started immediately when the field of transit spectroscopy with
{\it Spitzer} opened up in 2007. Near-IR spectroscopy with HST and
from the ground can be a powerful complement to these data because of
the clean spectral features at 1--1.6 $\mu$m
(Fig.~\ref{fig:earthspectrum}). Early detections of water in HD
189733b, HD 209458 and XO1b were not uniformly accepted
\citep[see][and chapter by \emph{Madhusudhan et al.} for
summaries]{Seager10,Tinetti12}. Part of the problem may stem from 
hazes or clouds that can affect the spectra shortward of 1.5
$\mu$m and hide water which is visible at longer wavelengths. Even for
the strongest cases for detection, it is difficult to retrieve an
accurate water abundance profile from these data, because of the low
$S/N$ and low spectral resolution \citep{Madhusudhan10}.  The advent
of a spatial scan mode has improved HST's ability to detect
exoplanetary water with WFC3 \citep[e.g.,][]{Deming13,Mandell13}. New
ground-based techniques are also yielding improved spectra
\citep{Birkby13}, but dramatic improvements in sensitivity and
wavelength range must await JWST. Until that time, only qualitative
conclusions can be drawn.

\begin{figure}[t]
\begin{centering}
\includegraphics[angle=0,width=0.4\textwidth]{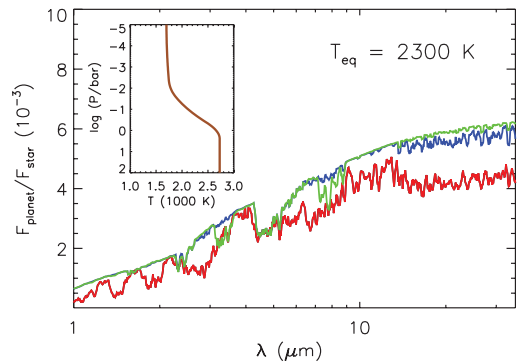}
\caption{Spectra of the hot Jupiters with C/O ratios varying from 0.54
  (solar, red) to 1 (blue) and 3 (green). Note the disappearance of
  the H$_2$O features at 1--3 $\mu$m as the C/O ratio increases.
  Figure from \citet{Madhusudhan11b}.}
\label{fig:hotjupiter}
\end{centering}
\end{figure}

\subsection{Gas-giant planets}

Models of hydrogen-dominated atmospheres of giant exoplanets
($T\approx 700-3000$~K; warm Neptune to hot Jupiter) solve for the
molecular abundances (i.e., ratios of number densities, also called
`mixing ratios') under the assumption of thermochemical
equilibrium. Non-equilibrium chemistry is important if transport
(e.g., vertical mixing) is rapid enough that material moves out of a
given temperature and pressure zone on a timescale short compared with
that needed to reach equilibrium. The CH$_4$/CO ratio is most affected
by such `disequilibrium' chemistry, especially in the upper
atmospheres. Since CO locks up some of the oxygen, this can also
affect the H$_2$O abundance.

The two main parameters determining the water abundance profiles are
the temperature-pressure combination and the C/O ratio. Giant planets
often show a temperature inversion in their atmospheres, with
temperature increasing with height into the stratosphere, which
affects both the abundances and the spectral appearance. However, for
the full range of pressures ($10^2$--$10^{-5}$ bar) and temperatures
(700--3000 K), water is present at abundances greater than $10^{-3}$
with respect to H$_2$ for a solar C/O abundance ratio of 0.54, i.e.,
most of the oxygen is locked up in H$_2$O. In contrast, when the C/O
ratio becomes close to unity or higher, the H$_2$O abundance can drop
by orders of magnitude especially at lower pressures and higher
temperatures, since the very stable CO molecule now locks up the bulk
of the oxygen \citep{Kuchner05,Madhusudhan11b}. CH$_4$ and other
hydrocarbons are enhanced as well. As noted previously, observational
evidence (not without controversy) for at least one case of such a
carbon-rich giant planet, WASP-12b, has been found
\citep{Madhusudhan11a}.  Hot Jupiters like WASP-12b are easier targets
for measuring C/O ratios, because their elevated temperature profiles
allow both water and carbon-bearing species to be measured without the
interference from condensation processes that hamper measurements in
Jupiter and Saturn in our own solar system, although cloud-forming
species more refractory than water ice can reduce spectral contrast
for some temperature ranges.  Figure \ref{fig:hotjupiter} illustrates
the changing spectral appearance with C/O ratio for a hot Jupiter
($T=2300$ K); for cooler planets ($<1000$~K) the changes are less
obvious.

What can cause exoplanetary atmospheres to have very different C/O
ratios from those found in the interstellar medium or even in
their parent star? If giant planets indeed form outside the snow line
at low temperatures through accretion of planetesimals, then the
composition of the ices is a key ingredient in setting the C/O
ratio. The volatiles (but not the rocky cores) vaporize when they
enter the envelope of the planet and are mixed in the atmosphere
during the homogenization process. As the giant planet migrates
through the disk, it encounters different conditions and thus
different ice compositions as a function of radius
\citep{Mousis09,Oberg11}. Specifically a giant planet formed outside
the CO snow line at 20--30 AU around a solar mass star may have a
higher C/O ratio than that formed inside the CO (but outside the
H$_2$O) snow line. 

The exact composition of these planetesimals depends on the adopted
model. The solar system community traditionally employs a model of the
solar nebula disk, which starts hot and then cools off with time
allowing various species to re-freeze. In this model, the abundance
ratios of the ices are set by their thermodynamic properties
(`condensation sequence') and the relative elemental abundances. For
solar abundances, water is trapped in various clathrate hydrates like
NH$_3$-H$_2$O and H$_2$S-5.75 H$_2$O between 5 and 20 AU. In a
carbon-rich case, no water ice is formed and all the oxygen is CO,
CO$_2$ and CH$_3$OH ice \citep{Madhusudhan11b}. The interstellar
astrochemistry community, on the other hand, treats the entire disk
with non-equilibrium chemistry, both with height and radius. Also, the
heritage of gas and ice from the protostellar stage into the disk can
be considered \citep[see
Fig.~\ref{fig:diskhistory}]{Aikawa99,Visser09,Aikawa12,Hincelin13}.  The
non-equilibrium models show that indeed the C/O ratio in ices can vary
depending on location and differ by factors of 2--3 from the overall
(stellar) abundances \citep{Oberg11}.

\subsection{Rocky planets, super-Earths}

The composition of Earth-like and super-Earth planets (up to 10
$M_{\Earth}$) is determined by similar thermodynamic arguments, with
the difference that most of the material stays in solid form and no
hydrogen-rich atmosphere is attracted. Well outside the snow line, at
large distances from the parent star, the planets are built up largely
from planetesimals that are half rock and half ice. If these planets
move inward, the water can become liquid, resulting in `ocean planets'
or `waterworlds', in which the entire surface of the planet is covered
with water \citep{Kuchner03,Leger04}. The oceans on such planets could
be hundreds of kilometers deep and their atmospheres are likely
thicker and warmer than on Earth because of the greenhouse effect of
water vapor. Based on their relatively low bulk densities (from the
mass-radius relation), the super-Earths GJ 1214b \citep{Charbonneau09}
and Kepler 22b \citep{Borucki12} are candidate ocean planets.

Further examination of terrestrial planet models shows that there
could be two types of water worlds: those which are true water rich in
their bulk composition and those which are mostly rocky but have a
significant fraction of their surface covered with water (Earth is in
the latter category) \citep{Kaltenegger13}.  Kepler 62e and f are
prototypes of water-rich planets within the habitable zone, a category
of planets that does not exist in our own solar system
\citep{Borucki13}. Computing the atmospheric composition of
terrestrial exoplanets is significantly more complex than that of
giant exoplanets and requires consideration of many additional
processes, including even plate tectonics \citep{Meadows11,Fortney13}.

As for the terrestrial planets in our own solar system, the presence
of one or more giant planet can strongly affect the amount of water on
exo-(super)-Earths (Fig.~\ref{fig:mixing}). 
A wide variety of dynamical and population synthesis models on
possible outcomes have been explored
\citep[e.g.][]{Raymond06b,Mandell07,Bond10,Ida04,Ida10,Alibert11,Mordasini09}. 

Water on a terrestrial planet in the habitable zone may be cycled many
times between the liquid oceans and the atmosphere through evaporation
and rain-out. However, only a very small fraction of water molecules
are destroyed or formed over the lifetime of a planet like Earth. The
total Earth hydrogen loss is estimated to be 3~kg\,s$^{-1}$. Even if
all the hydrogen comes from photodissociation of water, the water loss
would be 27 kg\,s$^{-1}$. Given the total mass of the hydrosphere of
$1.5\times10^{21}$~kg, it would take $1.8\times10^{12}$~yr to deplete
the water reservoir. The vast majority of the water bonds present
today were, therefore, formed by the chemistry that led to the bulk of
water in interstellar clouds and protoplanetary disks.

\section{\bf WATER TRAIL FROM CLOUDS TO PLANETS}

Here we summarize the key points and list some
outstanding questions.

\begin{dcl_itemize}

\item Water is formed on ice in dense molecular clouds. Some water is
  also formed in hot gas in shocks associated with protostars, but
  that water is largely lost to space.

\item Water stays mostly as ice during protostellar collapse and infall.
  Only a small fraction of the gas in the inner envelope is in a 'hot
  core' where the water vapor abundance is high due to ice
  sublimation.

\item Water enters the disk mostly as ice at large radii and is less
  affected by the accretion shock than previously thought.

\item Water vapor is found in three different reservoirs in
  protoplanetary disks: the inner gaseous reservoir, the outer icy belt
  and the hot surface layers. The latter two reservoirs have now been
  observed with {\it Spitzer} and {\it Herschel} and quantified.

\item Models suggest that the water that ends up in the planet and comet
  forming zones of disks is partly pristine ice and partly processed
  ice, i.e., ice that has at least once sublimated and recondensed
  when the material comes close to the young star.

\item Water ice promotes grain growth to larger sizes; water-coated
  grains grow rapidly to planetesimal sizes in dust traps that have
  now been observed.

\item Planetesimals inside the disk's snow line are expected to be
  dry, as found in our solar system. However, gaseous water can have
  very high abundances in the inner AU in the gas-rich phase.  Some
  grains may have chemically bound water to silicates at higher
  temperatures, but there is no evidence of hydrated silicates in
  meteorites nor in interstellar grains.

\item Dynamics affects what type of planetesimals are available for
  planet formation in a certain location. The presence and migration
  of giant planets can cause scattering of water-rich planetesimals
  from the outer disk into the inner dry zone.

\item Several comets have now been found with D/H ratios in water
  consistent with that of Earth's oceans. This helps to constrain
  models for the origin of water on the terrestrial planets, but does
  not yet give an unambiguous answer.

\item Both dry and wet formation scenarios for Earth are still open,
  although most arguments favor accretion of water-rich planetesimals
  from the asteroid belt during the late stages of terrestrial planet
  formation.  Delivery of water through bombardment to planetary
  surfaces in a `late veneer' can contribute as well but may not have
  been dominant for Earth, in contrast with Mars.

\item Jupiter may be poor in oxygen and water. If confirmed
  by the Juno mission to Jupiter, this may indicate a changing C/O
  ratio with disk radius. Evidence for this scenario comes from the
  detection of at least one carbon-rich exoplanet.

\item All of the processes and key parameters identified here should
  also hold for exo-planetary systems.

\end{dcl_itemize}

There are a number of open questions that remain.  A critical phase is
the feeding of material from the collapsing core onto the disk, and
the evolution of the young disk during the bulk of the phase of star
formation.  At present this phase has little observational
constraints.  In addition numerous models posit that viscous evolution
and the relative movement of the dust to the gas can have impact on
the overall water vapor evolution; this needs an observational basis.
Astronomical observations have been confined to water vapor and ice
emission from the disk surface but with the midplane hidden from view.
What does this surface reservoir tells us about forming planets, both
terrestrial and giant?  Is there a way to detect the midplane, perhaps
using HCO$^{+}$ (which is destroyed by water) as a probe of the
snowline?  However, this needs a source of ionization in the densest
parts of the disk which may or may not be present \citep{Cleeves13}.

The exploration of deuterium enrichments continues to hold promise
both in the solar system, with need for more information on the D/H
values for water in the outer asteroids and (main belt) comets.
It is also clear that we
know less about the bulk elemental abundance of the solar system's
largest reservoir of planetary material (i.e., Jupiter and Saturn)
than one might have assumed.  This must have 
implications for studies of extra-solar systems.  In terms of the
origin of the Earth's oceans matching the full range of geochemical
constraints remains difficult.  
    
Despite the uncertainties, there is a better understanding of
the water trail from the clouds to planets.  In this light, it is
fascinating to consider that most of the water molecules in Earth's
oceans and in our bodies may have been formed 4.6 billion years ago in
the cloud out of which our solar system formed.

\textbf{ Acknowledgments.} The authors thank many colleagues for
collaborations and input, and various funding agencies for support,
including NASA/JPL/Caltech. Figures by Magnus Persson, Ruud Visser,
Lars Kristensen and Davide Fedele are much appreciated.

\bigskip


\end{document}